\documentclass[preprint2]{emulateapj}

\newcommand{\Ka}{K$\alpha$}
\newcommand{\Ha}{H$\alpha$}
\newcommand{\NH}{$N_{\rm H}$}
\newcommand{\chisq}{$\chi ^2$/dof}
\newcommand{\lumi}{ergs~s$^{-1}$}
\newcommand{\flux}{ergs~cm$^{-2}$~s$^{-1}$}
\newcommand{\sbn}{ergs~cm$^{-2}$~s$^{-1}$~arcmin$^{-2}$}

\shorttitle{Evidence against Nonthermal X-Rays from N11 and N51D}
\shortauthors{Yamaguchi, Sawada, and Bamba}

\begin{document}

\title{Searching for Diffuse Nonthermal X-Rays from the Superbubbles \\
N11 and N51D in the Large Magellanic Cloud}

\author{H. Yamaguchi\altaffilmark{1}}
\email{hiroya@crab.riken.jp}
\author{M. Sawada\altaffilmark{2}}
\author{A. Bamba\altaffilmark{3,4}}

\altaffiltext{1}{RIKEN (The Institute of Physical and Chemical Research), 
  2-1 Hirosawa, Wako, Saitama 351-0198, Japan}
\altaffiltext{2}{Department of Physics, Kyoto University, 
  Kitashirakawa-oiwake-cho, Sakyo-ku, Kyoto 606-8502, Japan}
\altaffiltext{3}{Dublin Institute for Advanced Studies, 
  School of Cosmic Physics, 31 Fitzwilliam Place, Dublin 2, Ireland}
\altaffiltext{4}{Institute of Space and Astronautical Science, JAXA, 
  3-1-1 Yoshinodai, Sagamihara, Kanagawa 229-8510, Japan}

\begin{abstract}

We report on observations of the superbubbles (SBs) N11 and N51D in the 
Large Magellanic Cloud (LMC) with {\it Suzaku} and {\it XMM-Newton}. 
The interior of both SBs exhibits diffuse X-ray emission, which is well 
represented by thin thermal plasma models with a temperature of 0.2--0.3~keV. 
The presence of nonthermal emission, claimed in previous works, 
is much less evident in our careful investigation. 
The 3$\sigma$ upper limits of 2--10~keV flux are 
$3.6 \times 10^{-14}$~\flux\ and $4.7 \times 10^{-14}$~\flux\ 
for N11 and N51D, respectively. 
The previous claims of the detection of nonthermal emission are 
probably due to the inaccurate estimation of the non X-ray background. 
We conclude that no credible nonthermal emission has been 
detected from the SBs in the LMC, with the exception of 30~Dor~C.

\end{abstract}


\section{Introduction}
\label{sec:introduction}

\begin{table*}
\begin{center}
\caption{Log of observations appearing in this paper.
  \label{tab:log}}
\begin{tabular}{llccccc}
\tableline\tableline
 Target & Telescope/ & Observation & \multicolumn{2}{c}{Aim point} & 
    Observation & Effective  \\
 name   & Instrument & ID  & $\alpha$(J2000.0) & $\delta$(J2000.0) &
    start date & exposure (ks) \\
\tableline
 N11 & Suzaku/XIS  & 501091010 & $04^{\rm h}56^{\rm m}48^{\rm s}$ &
    $-66^{\circ}24'38''$ & 2006 Nov 7 & 30.5  \\
 ~ & XMM-Newton/EPIC  & 0109260201 & $04^{\rm h}55^{\rm m}58^{\rm s}$ &
    $-66^{\circ}25'53''$ & 2001 Sep 17 & 32.5 (MOS)/25.1 (pn)  \\
 N51D & Suzaku/XIS  & 804009010 & $05^{\rm h}25^{\rm m}37^{\rm s}$ &
    $-67^{\circ}29'57''$ & 2009 May 17 & 85.5/87.1/87.3\tablenotemark{a}  \\
 ~ & XMM-Newton/EPIC  & 0071940101 & $05^{\rm h}26^{\rm m}03^{\rm s}$ &
    $-67^{\circ}28'52''$ & 2001 Oct 31 & 30.5 (MOS)/23.1 (pn)  \\
 NEP & Suzaku/XIS  & 500026010 & $18^{\rm h}11^{\rm m}18^{\rm s}$ &
    $+66^{\circ}00'45''$ & 2006 Feb 10 & 88.5  \\
  \tableline
\end{tabular}
\tablecomments{
$^{a}$The exposure times of the XIS0/1/3.
The XIS2 was out of operation during this observation. 
}
\end{center}
\end{table*}

Massive stars are often found clustered in OB associations. 
Their high-speed stellar winds and the blast wave from core-collapsed 
supernova (SN) explosions collectively sweep up the ambient 
interstellar medium (ISM) to generate large (10--100-pc-scale) 
shell-like structures known as superbubbles (SBs) 
(e.g., McCray \& Snow 1979; Mac Low \& McCray 1988). 
A tenuous cavity created inside an SB shell allows the blast wave of 
a subsequent supernova remnant (SNR) to propagate rapidly without 
deceleration for a long time. 
Furthermore, wind-wind and/or shock-cloud interactions, expected inside 
an SB, can maintain significant turbulence and magnetic inhomogeneities. 
Thus, SBs are believed to be very efficient site for the acceleration 
of relativistic cosmic rays 
(e.g., Bykov \& Fleishman 1992; Parizot et al.\ 2004).

The Large Magellanic Cloud (LMC) is particularly suitable for the 
study of SBs, because of its low foreground extinction \citep{dl} 
and its known distance of $\sim$50~kpc with a small uncertainty 
(e.g., Persson et al.\ 2004). 
Using {\it ROSAT}, Chu \& Mac Low (1990) and \cite{dunne01} 
systematically studied several SBs in the LMC. They found that 
the cavities inside the SBs exhibit diffuse X-ray emissions from 
rarefied hot gas with a temperature on the order of 0.1~keV. 
They also showed that the X-ray surface brightnesses 
are significantly higher than those expected from the standard 
pressure-driven SB model (Weaver et al.\ 1977). 
This suggests the presence of interior SNRs colliding with 
the inner walls of the SB shell.

The detection of nonthermal X-rays has been reported in three LMC SBs, 
30~Dor~C \citep{bamba04}, N11 \citep{m09}, and N51D \citep{c04}. 
30~Dor~C exhibits the nonthermal emission with a clear shell-like 
structure. The spectra obtained with {\it Chandra} and {\it XMM-Newton} 
were represented by a power-law with a photon index of $\Gamma$ = 2--3 
\citep{bamba04}, typical for synchrotron X-rays from nonthermal SNRs. 
Using {\it Suzaku}, \cite{yama09} discovered the spectral cutoff 
in the power-law emission. They concluded that the synchrotron X-rays 
originate from the rapidly-expanding shell of the SNR with an age of 
(4--20) $\times 10^3$~yr.

On the other hand, the validities of the other two results may be 
doubtful. \cite{m09} have recently observed N11 with {\it Suzaku} 
and have reported the detection of nonthermal X-rays above 2~keV. 
However, the claimed flux is comparable to that of 30~Dor~C despite 
the fact that no nonthermal component was detected by the previous 
{\it XMM-Newton} observation \citep{naze04}. 
Moreover, a morphology of the emission has been unclear.  
The {\it Suzaku} data should be therefore carefully reanalyzed.

The nonthermal emission in N51D is claimed to be distributed entirely 
in the SB cavity with a centrally peaked morphology in contrast to 
30~Dor~C \citep{c04}. The photon index of $\Gamma \sim 1.3$ is 
much lower than those expected for synchrotron X-rays. 
On the other hand, in the preceding work by \cite{bomans03}, 
the same {\it XMM-Newton} spectrum was well represented by a thermal 
plasma model without any additional nonthermal components. 
Thus, the existence of nonthermal emission in N51D is still controversial. 
Motivated by this issue, we have obtained a long-time 
exposure of this SB with {\it Suzaku}.

The aim of this paper is to investigate for the existence of diffuse hard 
X-ray emission. Therefore, other properties of SBs and their stellar 
contents are not focused on. 
For the analyses, the {\it Suzaku} data are mainly utilized, 
since these provide a low and stable non-X-ray background (NXB), 
particularly for diffuse sources (Mitsuda et al.\ 2007; Tawa et al.\ 2008). 
In addition, we also refer to {\it XMM-Newton} archival data to 
compensate the limited angular resolution of {\it Suzaku}.

First, we briefly introduce the characteristics of N11 
(\S\ref{ssec:intro_n11}) and N51D (\S\ref{ssec:intro_n51d}). 
In Section~\ref{sec:observation}, we outline our observation and data 
reduction procedure. The results for N11 and N51D are presented 
in Sections~\ref{sec:n11} and \ref{sec:n51d}, respectively. 
We discuss our results in Section~\ref{sec:discussion}, 
and give a summary in Section~\ref{sec:summary}. 
The errors quoted in the text and tables refer to the 90\% confidence 
intervals for one parameter, and the error bars in the figures 
are for 1$\sigma$ confidence, unless otherwise stated.

\subsection{Characteristics of N11}
\label{ssec:intro_n11}

N11 (DEM L~34), located at the northwest edge of the LMC \citep{henize56}, 
contains four OB associations, LH9, LH10, LH13, and LH14 \citep{lh}. 
The \Ha\ emission exhibits a complex structure consisting of several 
shells and filaments (e.g., Mac~Low et al.\ 1998). 
The largest \Ha\ shell with a diameter of $\sim$120~pc is situated 
around the central cluster LH9. 
The age of LH9 is estimated to be $\sim$3.5~Myr (Walborn et al.\ 1999), 
the oldest among the four clusters. It is suggested that the stellar 
activity of LH9 triggered a starburst at the periphery, 
leading to the formation of the other OB associations 
(e.g., Walborn \& Parker 1992; Hatano et al.\ 2006).

\subsection{Characteristics of N51D}
\label{ssec:intro_n51d}

N51D (DEM L~192) is located at the southeast corner of the 
supergiant shell LMC4. Two OB associations, LH51 and LH54 \citep{lh}, 
are surrounded by a filamentary SB shell with a size of about 
140~pc $\times$ 120~pc. Both the OB clusters are estimated 
to have an age of $\sim$3~Myr (Oey \& Smedley 1998).

\section{Observation and Data Reduction}
\label{sec:observation}

\subsection{Suzaku}
\label{ssec:obs_s}

The SBs N11 and N51D were observed with the {\it Suzaku} 
X-ray Imaging Spectrometer (XIS: Koyama et al.\ 2007) on 
2006 November 7 and 2009 May 17, respectively. The observation 
IDs and aim points are summarized in Table~\ref{tab:log}. 
The XIS consists of four X-ray charge coupled devices (CCDs). 
Three of them (XIS0, XIS2, and XIS3) are front-illuminated (FI) and 
the other (XIS1) is back-illuminated (BI). 
XIS0, 2, and 3 have a lower and more stable background level for 
extended emission, while the latter has superior sensitivity in 
the soft ($\lesssim$1.0~keV) X-ray band with a significantly improved 
energy resolution compared with previous X-ray imaging sensors. 
Combined with X-Ray Telescopes (XRTs; Serlemitsos et al.\ 2007), 
the field of view (FOV) of the XIS covers a $\sim$$18'$$\times$$18'$ 
region with a half-power diameter (HPD) of $\sim$$2'$.
During both the observations, the XIS was operated in the normal 
full-frame clocking mode with spaced-row charge injection technique. 
However, the XIS2 was out of operation during the observation of N51D 
due to damage, possibly caused by the impact of a micro meteorite.

We reduced the data using the standard tools of HEADAS version 6.7. 
We first employed the revision 2.0 and 2.4 data products for N11 and 
N51D, respectively. Both were reprocessed using the \texttt{xispi} 
software and the \texttt{makepi} files released on 2009 August 13. 
These files include the latest calibration results for the charge 
transfer inefficiency and gain. 
We cleaned the reprocessed data in accordance with the standard screening
criteria\footnote{http://heasarc.nasa.gov/docs/suzaku/processing/criteria\_xis.html} and removed flickering pixels. 
We also excluded time intervals that suffered from telemetry saturation. 
After the filtering, the effective exposure times were obtained to be 
$\sim$31~ks and $\sim$86~ks for the observations of N11 and N51D, 
respectively. 
Only grade 0, 2, 3, 4, and 6 events were used in the following analysis.

\subsection{XMM-Newton}
\label{ssec:obs_x}

Detailed information on the {\it XMM-Newton} observations of 
both SBs is given in Table~\ref{tab:log}. 
The European Photon Imaging Camera (EPIC), aboard {\it XMM-Newton}, has 
two Metal Oxide Semi conductor (MOS) CCD arrays (Turner et al.\ 2001)
and one pn CCD array  (Str{\" u}der et al. 2001). 
The latter has a higher quantum efficiency than the other CCD arrays. 
All three cameras have a larger FOV ($\sim$$30'$-diameter circle) than 
the XIS with a much better HPD of $\sim$$6''$ at the optical axis. 
This makes observations with the EPIC highly sensitive to 
point-like sources. 
During both the observations, the MOS and pn CCD arrays were operated in 
prime full window mode and extended prime full window mode, respectively. 
Thick filters were used in the observation of N11, while 
thin ones were used in the observation of N51D.

All the data were processed using version 9.0.0 of the XMM Science 
Analysis Software (SAS) with the latest calibration files. 
For the MOS, we selected X-ray events with patterns 0--12, 
which were passed through the \#XMMEA\_EM filter. For the pn, 
only events with patterns 0--4 and flag=0 were extracted. 
Light curves in the 10--15~keV band, where the effective area of the 
telescopes rapidly decreases, were accumulated from the whole FOV. 
To minimize any uncertainty due to the NXB, we removed the periods 
when the count rate was larger than 0.3~counts~s$^{-1}$ (MOS) 
and 1.0~counts~s$^{-1}$ (pn). 
The resultant exposure times are given in Table~\ref{tab:log}.

\section{Results for N11}
\label{sec:n11}

\subsection{Images and region selection}
\label{ssec:n11_image}

\begin{figure*}
  \begin{center}
  \includegraphics[scale=.34]{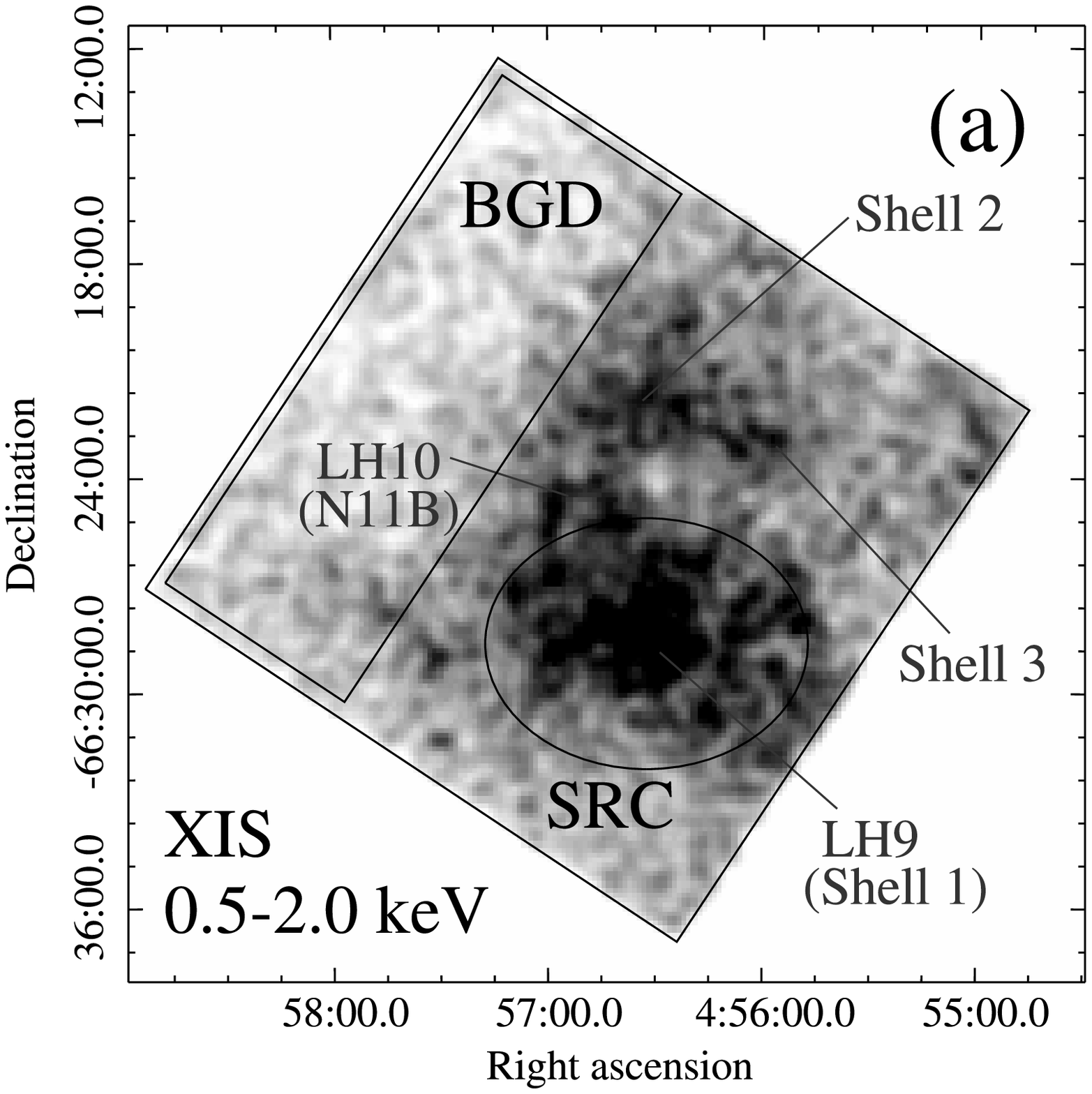}
  \includegraphics[scale=.34]{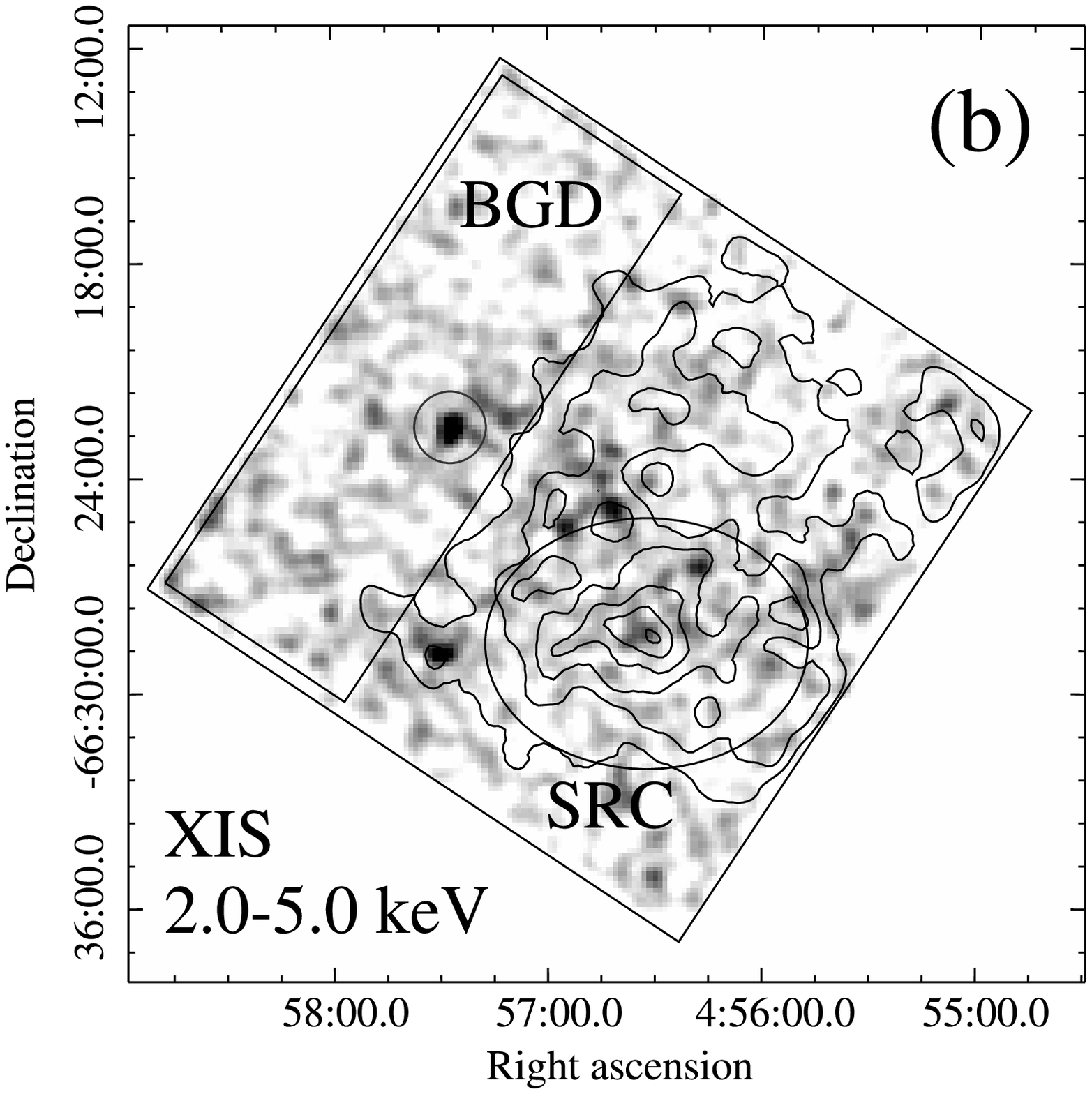}
  \includegraphics[scale=.31]{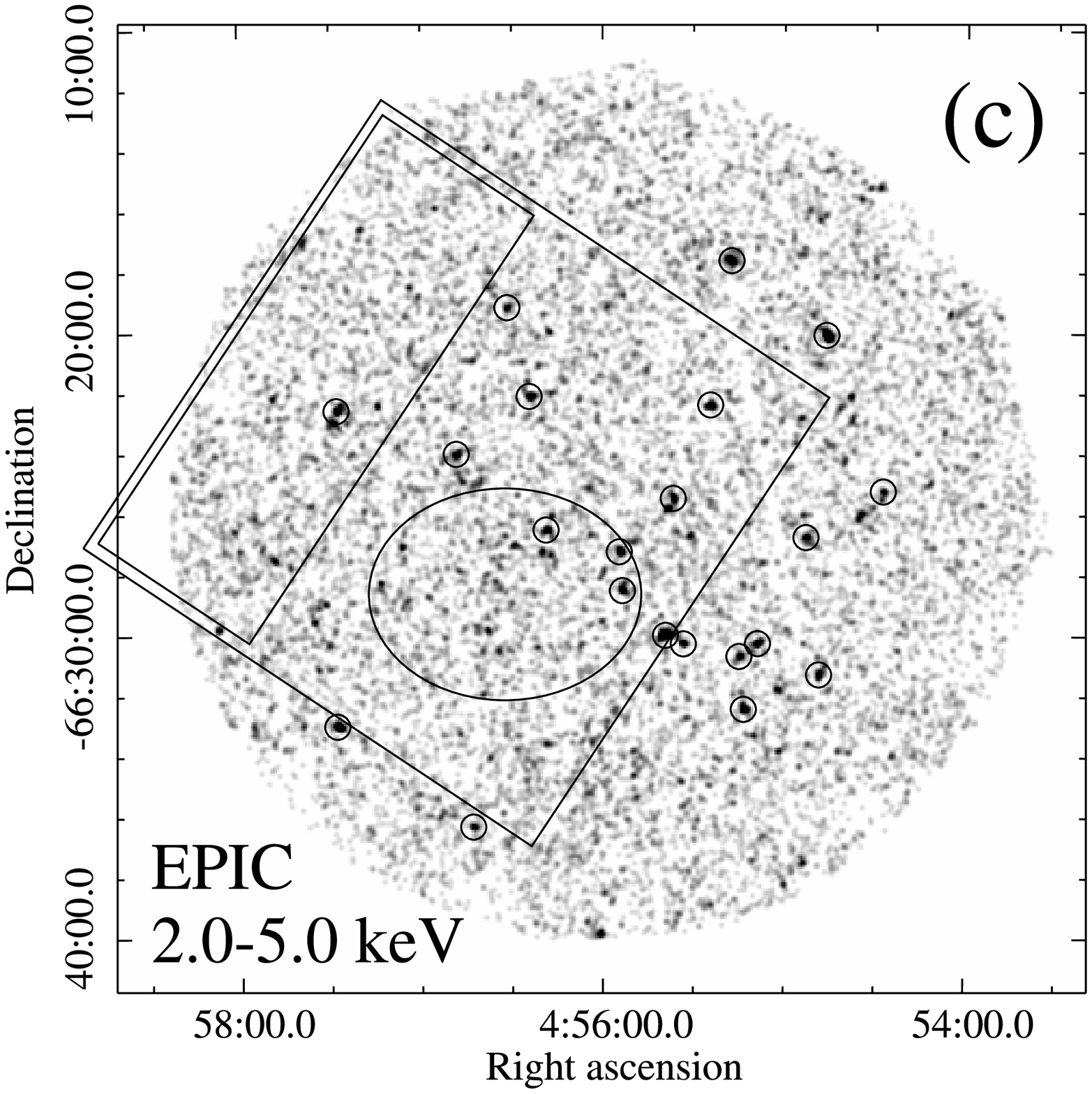}
\caption{Images of N11 SB region, displayed on the J2000.0 
  coordinates. Panels (a) and (b) are {\it Suzaku} XIS images in the 
  0.5--2.0~keV and 2.0--5.0~keV bands, respectively. The latter is 
  overlaid with contours of the former energy band smoothed with 
  a Gaussian kernel of $\sigma = 42''$. The ellipse and rectangle 
  indicate the regions of source (SRC) and background (BGD) spectra, 
  respectively. The small region shown as the gray circle in the 
  panel (b) is excluded from the BGD spectrum. 
  A panel (c) is {\it XMM-Newton} EPIC-MOS image in the 2.0--5.0~keV 
  band. The FOV of the {\it Suzaku} XIS and the regions for the 
  spectral analysis are shown. 
  Detected point sources are expressed as the small circles. 
  Note that Shells~1, 2, and 3, labeled in the panel (a), are 
  named by \cite{mac98}.
  \label{fig:n11_image}}
  \end{center}
\end{figure*}

Figures~\ref{fig:n11_image}(a) and \ref{fig:n11_image}(b) show XIS images 
of the N11 region in the soft (0.5--2.0~keV) and hard (2.0--5.0~keV) 
X-ray bands, respectively. Data from the four detectors were combined. 
The images were binned by a factor of 8 and smoothed with a Gaussian 
kernel of $\sigma = 25''$. 
Extended emission can be clearly seen in the soft-band image, 
while no significant diffuse structure was found above 2.0~keV.

As already mentioned in previous works (e.g., Naz{\'e} et al.\ 2004), 
the soft emission peaks near the dense stellar cluster HD~32228 and is 
widely associated with the OB association LH9. This part is confined by 
the largest \Ha\ shell of the N11 complex \citep{mac98}, 
and thus it is considered that the soft X-rays originate from hot 
shocked gas within the SB shell that is blown by the cluster of stars. 
The emission is further extended to the north of LH9, the HII region 
N11B, and shells~2 and 3 (see also Figure~3 of Mac~Low et al.\ 1998). 
The X-rays from N11B are probably powered by stellar winds from 
the central young OB stars of LH10. Shells~2 and 3 lie to the north and 
northwest of N11B, respectively. These X-ray emissions with relatively 
low surface brightnesses are confined by the faint \Ha\ filaments.

\cite{m09} claimed that nonthermal X-rays had been detected from 
the region around LH9. Therefore, we concentrate on the spectral 
analysis of the same region, an ellipse containing the entire SB 
around LH9 with major and minor radii of $4.\!'5$ and $3.\!'5$, 
respectively---hereafter, ``SRC''. 
The background (BGD) for the SRC spectra was taken from an off-source 
region, the rectangle shown in Figure~\ref{fig:n11_image}, but the CCD 
corners were excluded to remove X-rays from the calibration sources.

A high-resolution EPIC image in the hard band is shown in 
Figure~\ref{fig:n11_image}(c). This image was made by merging two MOS 
images binned by a factor of 32 and smoothed with a Gaussian profile of 
$\sigma = 9.\!''6$. Similarly to the XIS image for the same energies 
(Figure~\ref{fig:n11_image}b), no extended emission was observed in 
the EPIC FOV. 
Point source extraction was performed using the \texttt{ewavelet} task of 
SAS with a detection threshold of 5$\sigma$. The positions of the 
detected sources are indicated with small circles in the figure. 
Three discrete sources are located in the SRC region, 
while one is located in the BGD region. The latter, at (R.A., Decl.) = 
(4$^{\rm h}$57$^{\rm m}$28$^{\rm s}$, --66$^{\circ}$22$'$40$''$), 
is known to be associated with the cluster HD~268743, or Sk--66$^{\circ}$41 
\citep{naze04}. Since this source is also clearly visible in the XIS image, 
we excluded it from the BGD region for spectral analysis.

\begin{figure}
  \begin{center}
    \rotatebox{270}{
    \includegraphics[scale=.31]{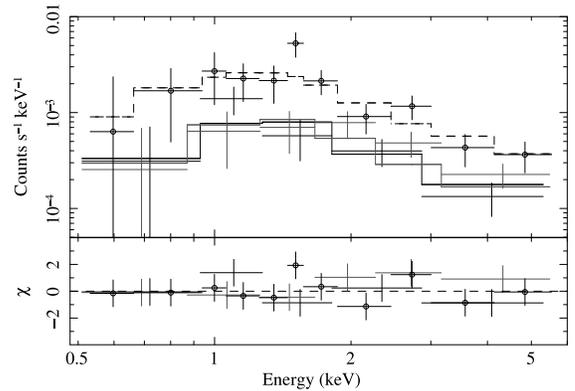}
    }
    \caption{Integrated EPIC spectra of three point sources 
      included within the N11 SRC region. 
      MOS1 (black), MOS2 (gray), and pn (black with circles) 
      spectra are simultaneously fitted with an absorbed power-law 
      model shown with solid and dashed lines. 
  \label{fig:n11_ptsrc}}
  \end{center}
\end{figure}

\begin{table*}
\begin{center}
\caption{2--10~keV surfaces brightnesses of N11 and NEP regions.
  \label{tab:sbn_n11}}
\begin{tabular}{lccc}
  \tableline\tableline
  ~ & Solid angle & Flux [2--10~keV] & Surface brightness [2--10~keV] \\
  ~ & (arcmin$^2$) & ($\times 10^{-13}$~\flux) & ($\times 10^{-15}$~\sbn) \\
  \tableline
  SRC    &  49.4    & 2.2 (2.0--2.4)  &  4.4 (4.0--4.8) \\
  ~~(point sources) & ~ & 0.36 (0.31--0.41)  &  --- \\
  ~~(subtracted)    & ~ & 1.8 (1.6--2.0)     &  3.6 (3.2--4.1) \\
  BGD    &  70.4    & 2.8 (2.5--3.1)  &  4.0 (3.6--4.4) \\
  NEP    & 113.1    & 5.0 (4.8--5.1)  &  4.4 (4.2--4.5) \\
  \tableline
\end{tabular}
%% Any table notes must follow the \end{tabular} command.
\tablecomments{
  The NXB is subtracted from each region, but the CXB is still included. 
}
\end{center}
\end{table*}

\begin{figure*}
  \begin{center}
    \rotatebox{270}{
    \includegraphics[scale=.31]{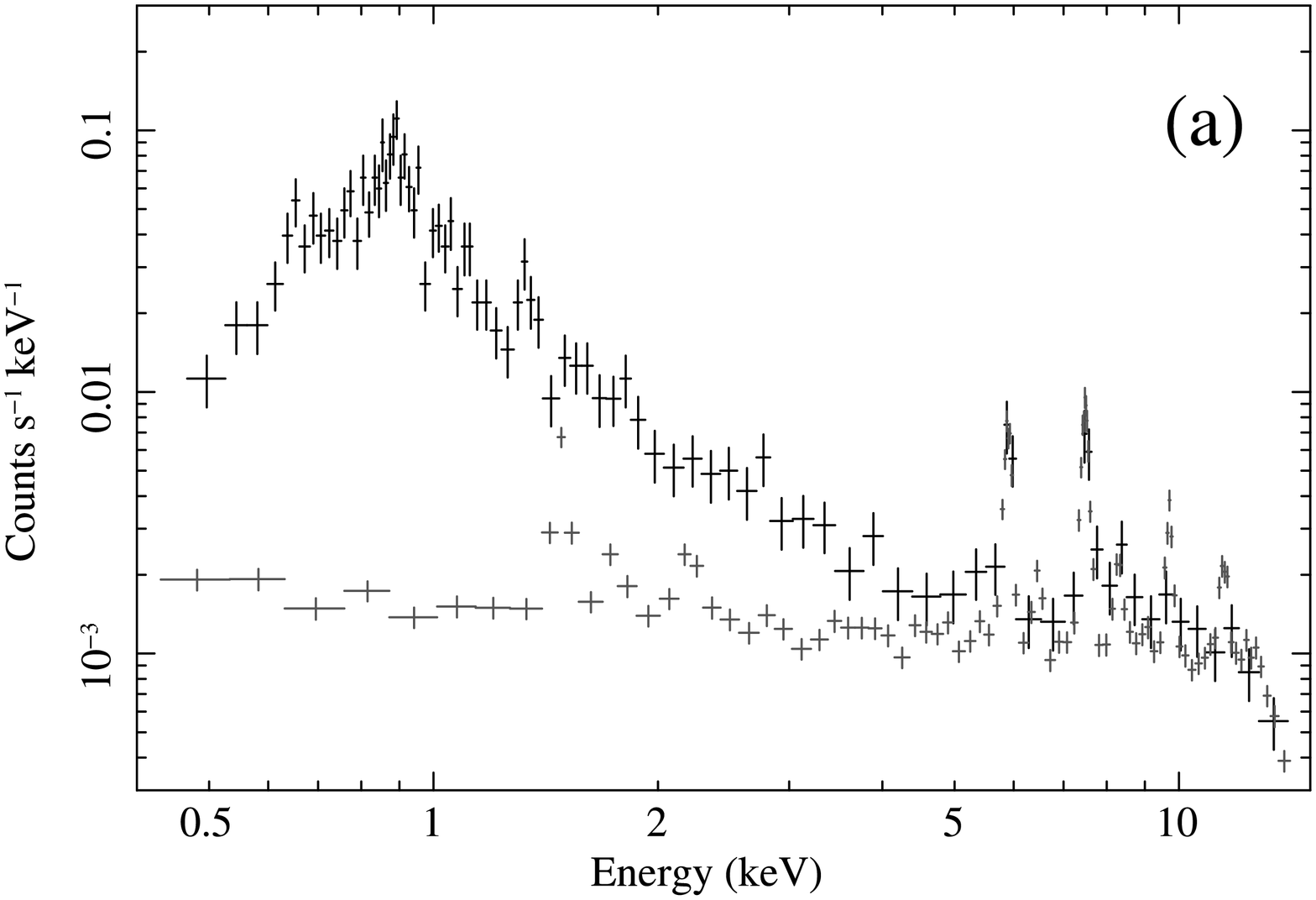}
    }
    \rotatebox{270}{
    \includegraphics[scale=.31]{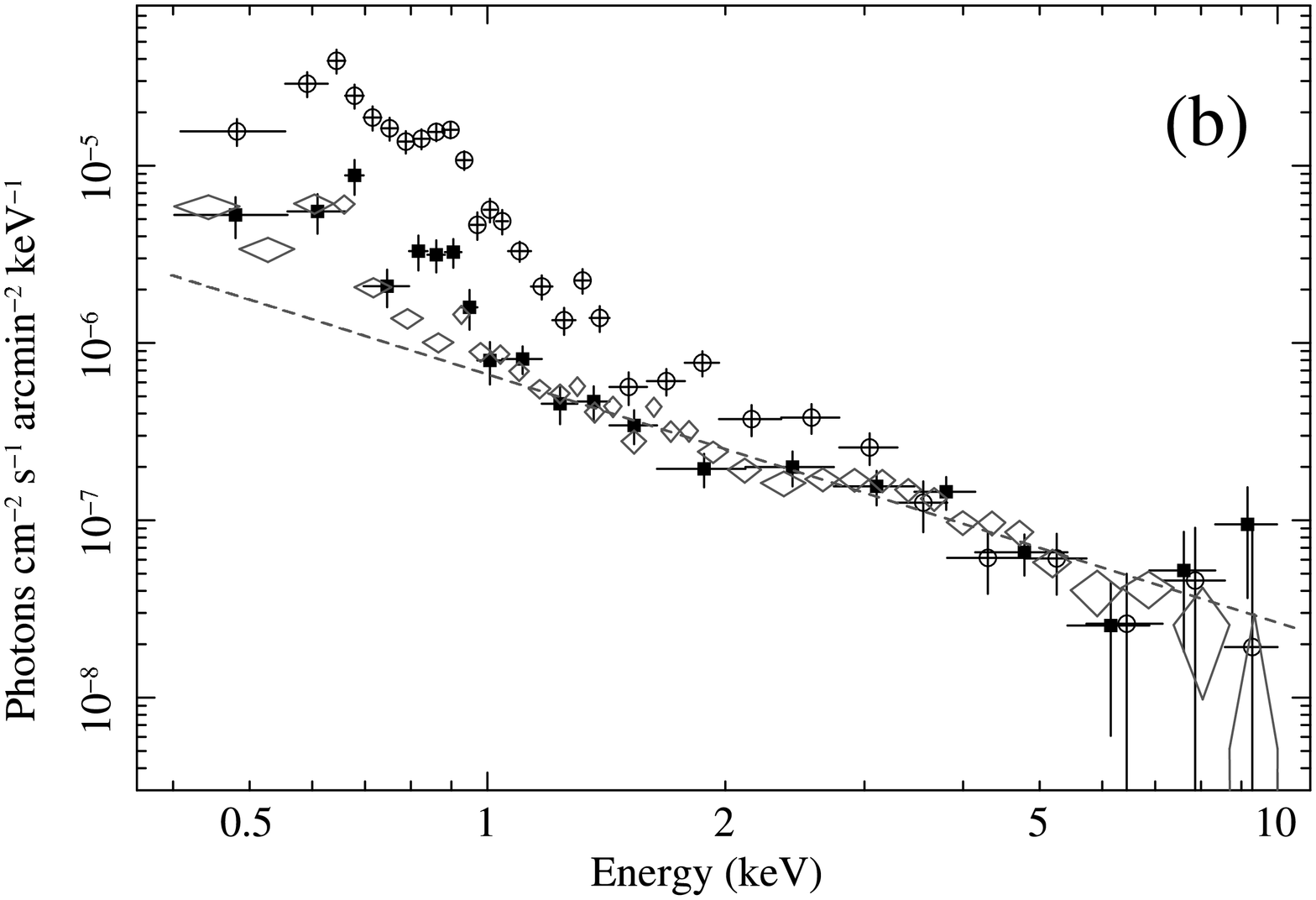}
    }
    \caption{(a) XIS0 raw spectrum of the N11 SRC region (black) 
      and NXB spectrum generated by 
      \texttt{xisnxbgen} software (gray). 
      (b) NXB-subtracted photon spectra of the SRC (open circle) and 
      BGD (filled square) regions, normalized by each solid angle. 
      For comparison, the same spectrum from the NEP are shown with 
      gray diamonds. The NEP spectrum was taken from an on-axis 
      circular region with a radius of 6~arcmin. 
      The dashed line indicates a power-law with photon index of 1.4, 
      a typical value for the cosmic X-ray background. 
  \label{fig:n11_compare}}
  \end{center}
\end{figure*}

\subsection{XMM-Newton spectra of point-like sources}
\label{ssec:n11_xmm}

Here we analyze the EPIC data of the resolved point sources to take 
into account their contribution to the hard X-ray flux from the SRC 
region. Figure~\ref{fig:n11_ptsrc} shows the background-subtracted 
EPIC spectra, where the three point sources in the SRC region 
are merged to improve the statistics. 
The spectrum of each point source was extracted from the circular region 
with a radius of $15''$, while the background was taken from an annulus 
with inner and outer radii of $15''$ and $60''$, respectively.

The spectra were phenomenologically fitted with an absorbed 
power-law model. The absorption cross sections were taken from 
Morrison \& McCammon (1983). The elemental abundances were assumed 
to be the values in \cite{angr}. 
The fit was acceptable (\chisq\ = 15/19) with 
\NH\ = 2.9 (1.1--7.8) $\times 10^{21}$~cm$^{-2}$ and 
$\Gamma$ = 1.5 (1.1--2.2). 
The observed flux in the 2--10~keV band was obtained to be 
3.6 (3.1--4.1) $\times 10^{-14}$~\flux.

\subsection{Suzaku spectra of extended emission}
\label{ssec:n11_suzaku}

Since the emission from the SRC region is faint, background reduction 
should be carefully performed to obtain an accurate result. 
We first subtract the NXB from the SRC and BGD raw spectra, 
and examine the difference in the hard X-ray fluxes between 
them (Section~\ref{sssec:n11_suzaku1}). 
Then, we subtract the other background components, such as the cosmic 
X-ray background (CXB) and local emission, and perform model fittings 
to the SRC spectrum (Section~\ref{sssec:n11_suzaku2}).

\subsubsection{Search for hard X-rays}
\label{sssec:n11_suzaku1}

Figure~\ref{fig:n11_compare}(a) shows the XIS0 raw spectrum 
of the SRC region. Instrumental emission lines of Mn-\Ka, 
Ni-\Ka\ and K$\beta$, and Au-L$\alpha$ can be observed at $\sim$5.9~keV, 
$\sim$7.5~keV, $\sim$8.3~keV, and $\sim$9.7~keV, respectively. 
Using \texttt{xisnxbgen} software, we constructed the NXB spectrum from 
the night-Earth database. This was extracted from the same detector 
coordinates to the SRC region and sorted with the geomagnetic cut off 
rigidity for reproducibility. 
The intensities of the Ni-\Ka\ lines were found to be  
1.6 (1.1--2.1) $\times 10^{-3}$~counts~s$^{-1}$ and 
1.6 (1.5--1.7) $\times 10^{-3}$~counts~s$^{-1}$ 
for the raw SRC and NXB spectra, respectively. 
Since these are highly consistent with each other, we can consider 
that the NXB spectrum was correctly reproduced. The NXB for the BGD 
spectrum was also successfully constructed by the same method.

After the NXB subtraction, the surface brightness spectra of the SRC 
and BGD regions were obtained as shown in Figure~\ref{fig:n11_compare}(b) 
(shown as circles and squares, respectively). 
For comparison, we also present the spectrum of another source-free 
region (shown as gray diamonds), taken from the observation of 
the north ecliptic pole (NEP). 
All the spectra are similar to each other at energies above $\sim$3~keV, 
suggesting that the hard X-rays from the SRC region are dominated by 
the CXB. 
Table~\ref{tab:sbn_n11} gives the 2--10~keV surface brightness of 
each region, determined using the data of all four XIS. 
No excess SRC flux was found.

\subsubsection{Spectral fitting}
\label{sssec:n11_suzaku2}

\begin{figure}
  \begin{center}
    \rotatebox{270}{
    \includegraphics[scale=.32]{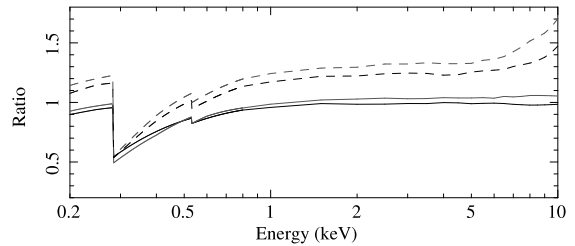}
    }
    \caption{Effective-area ratios between the SRC and BGD regions, 
      used for the vignetting correction. 
      Solid black, solid gray, dashed black, and dashed gray represent 
      XRT0, XRT1, XRT2, and XRT3, respectively. 
  \label{fig:n11_ratio}}
  \end{center}
\end{figure}

\begin{figure}
  \begin{center}
    \rotatebox{270}{
    \includegraphics[scale=.31]{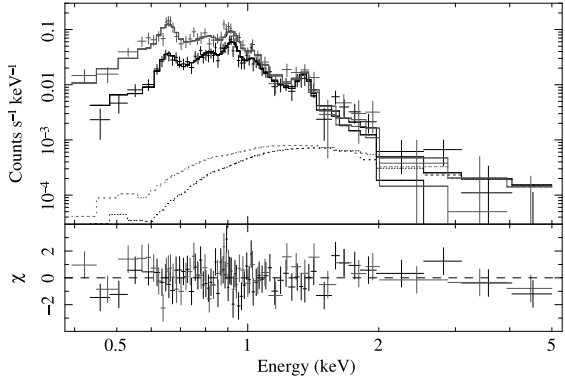}
    }
    \caption{Background-subtracted XIS spectra of N11, fitted 
      with an APEC model (solid lines). Black and gray represent 
      FI and BI, respectively. Contribution of the point sources 
      determined from the EPIC spectra is given with the dotted lines. 
  \label{fig:n11_diffuse}}
  \end{center}
\end{figure}

\begin{table}
\begin{center}
\caption{Best-fit spectral parameters for N11.
  \label{tab:fit_n11}}
\begin{tabular}{lc}
\tableline\tableline
  Parameter & Value  \\
  \tableline
  $N_{\rm H}^{\rm G}$ (cm$^{-2}$)\tablenotemark{a} & 
      $4.3 \times 10^{20}$ (fixed)   \\
  $N_{\rm H}^{\rm LMC}$ (cm$^{-2}$)\tablenotemark{a} & 
      $< 2.3 \times 10^{21}$   \\
  $kT_e$ (keV) & 0.30 (0.28--0.32) \\
  O  (solar) & 0.26 (0.15--0.47)   \\
  Ne (solar) & 0.39 (0.20--0.47)   \\
  Mg (solar) & 0.43 (0.22--0.89)   \\
  Fe (solar) & 0.12 (0.08--0.22)  \\
  $n_e n_p V$ (cm$^{-3}$) & 5.4 (2.6--19) $\times ~10^{58}$ \\
  $F_{\rm X}^{\rm obs}$~(\flux)\tablenotemark{b} & $4.9\times 10^{-13}$  \\
  $L_{\rm X}$~(\lumi)\tablenotemark{b} & $2.2\times 10^{35}$  \\
  \tableline
  \chisq & 99/105 = 0.94  \\
  \tableline
\end{tabular}
\tablecomments{
  $^{a}$Absorption column in the Galaxy and LMC, where the solar and 
  LMC abundances are respectively assumed. \ 
  $^{b}$Observed flux and luminosity in the 0.5--2.0~keV band.
}
\end{center}
\end{table}

The spectrum of the BGD region was subtracted from the SRC spectrum. 
However, since the vignetting of the {\it Suzaku} XRT significantly 
decreases the effective area at large off-axis positions 
(Serlemitsos et al.\ 2007), the BGD spectrum $I_{\rm B}(E)$
was corrected before the subtraction as 
\begin{equation}
  I_{\rm B}^{\prime}(E) = I_{\rm B}(E) \cdot 
  \frac{\Omega_{\rm S}}{\Omega_{\rm B}} \cdot f(E)~,
\end{equation}
where ${\Omega_{\rm S}}$ and ${\Omega_{\rm B}}$ are the solid angles 
of the SRC and BGD regions. The correction factor $f(E)$ is the 
energy-dependent effective-area ratio between the SRC and BGD regions. 
Using \texttt{xissim} software, we estimated the $f(E)$ values for 
each XRT as given in Figure~\ref{fig:n11_ratio}. 
This estimation also takes into account the contaminating material built 
up on the optical blocking filters of the XIS (Koyama et al.\ 2007). 
The differences in the results among the telescopes are mainly due to 
the scattering of the optical axes of the XRTs 
(see Figure~8 of Serlemitsos et al.\ 2007).

The BGD-subtracted SRC spectra are shown in Figure~\ref{fig:n11_diffuse}. 
The data from three FIs were combined because their response 
characteristics are almost identical. 
No significant signal was detected in the energy band above $\sim$3~keV, 
which is consistent with the result in Section~\ref{sssec:n11_suzaku1}. 
K-shell emission lines of O~VIII ($\sim$0.65~keV), Ne~IX ($\sim$0.91~keV), 
and Mg~XI ($\sim$1.34~keV) were clearly separate, indicating a thermal 
origin of the soft X-rays. Thus, we fitted the spectra with 
an optically thin thermal plasma model (APEC: Smith et al.\ 2001). 
The electron temperature ($kT_e$) and emission measure ($n_e n_p V$, 
where $n_e$, $n_p$, and $V$ are the electron and proton densities, and 
the emitting volume, respectively) were treated as free parameters. 
The elemental abundances relative to solar values \citep{angr}
were fixed to the mean LMC values of \cite{rd}, but the Mg and Si 
abundances were assumed to be those of Hughes et al.\ (1998) 
since the values from \cite{rd} were more uncertain. 
The interstellar extinction in the Galaxy and LMC were separately 
considered. The Galactic absorption column density with the solar 
abundances was fixed to be 
$N_{\rm H}^{\rm G}$ = $4.3\times 10^{20}$~cm$^{-2}$ \citep{dl}. 
The other component ($N_{\rm H}^{\rm LMC}$) was a free parameter, 
with the assumption of the LMC metal abundances. 
In addition, the point source model derived in Section~\ref{ssec:n11_xmm} 
was given as a fixed component (with an independent absorption column). 
This model gave a best fit with $kT_e$ = 0.18 (0.17--0.19)~keV and 
\chisq\ = 144/109.

We next allowed the abundances of O, Ne, Mg, and Fe to vary freely 
(by using a VAPEC model). Then, we obtained a significantly improved 
fit with \chisq\ = 99/105. 
The best-fit parameters are given in Table~\ref{tab:fit_n11}. 
The Fe abundance was found to be slightly lower than the mean LMC 
value ($\sim$0.3~solar: Russell \& Dopita 1992; Hughes et al.\ 1998). 
The observed flux of $4.9 \times 10^{-13}$~\flux\ 
(in the 0.5--2.0~keV band) is comparable with the result of \cite{naze04} 
but inconsistent with that of \cite{m09} ($\sim$$2.5 \times 10^{-13}$~\flux), 
although the same data were analyzed.

The derived electron temperature was slightly higher than the 
values in previous reports ($\sim$0.2~keV: Dunne et al.\ 2001; 
Naz{\'e} et al.\ 2004; Maddox et al.\ 2009). 
We found, however, that the fit with a fixed $kT_e$ of 0.2~keV 
also yielded an acceptable \chisq\ value of 111/106. 
This temperature difference does not severely affect 
the flux in the hard ($>$2~keV) X-ray band.

\section{Results for N51D}
\label{sec:n51d}

\subsection{Images and region selection}
\label{ssec:n51d_image}

\begin{figure*}
  \begin{center}
  \includegraphics[scale=.34]{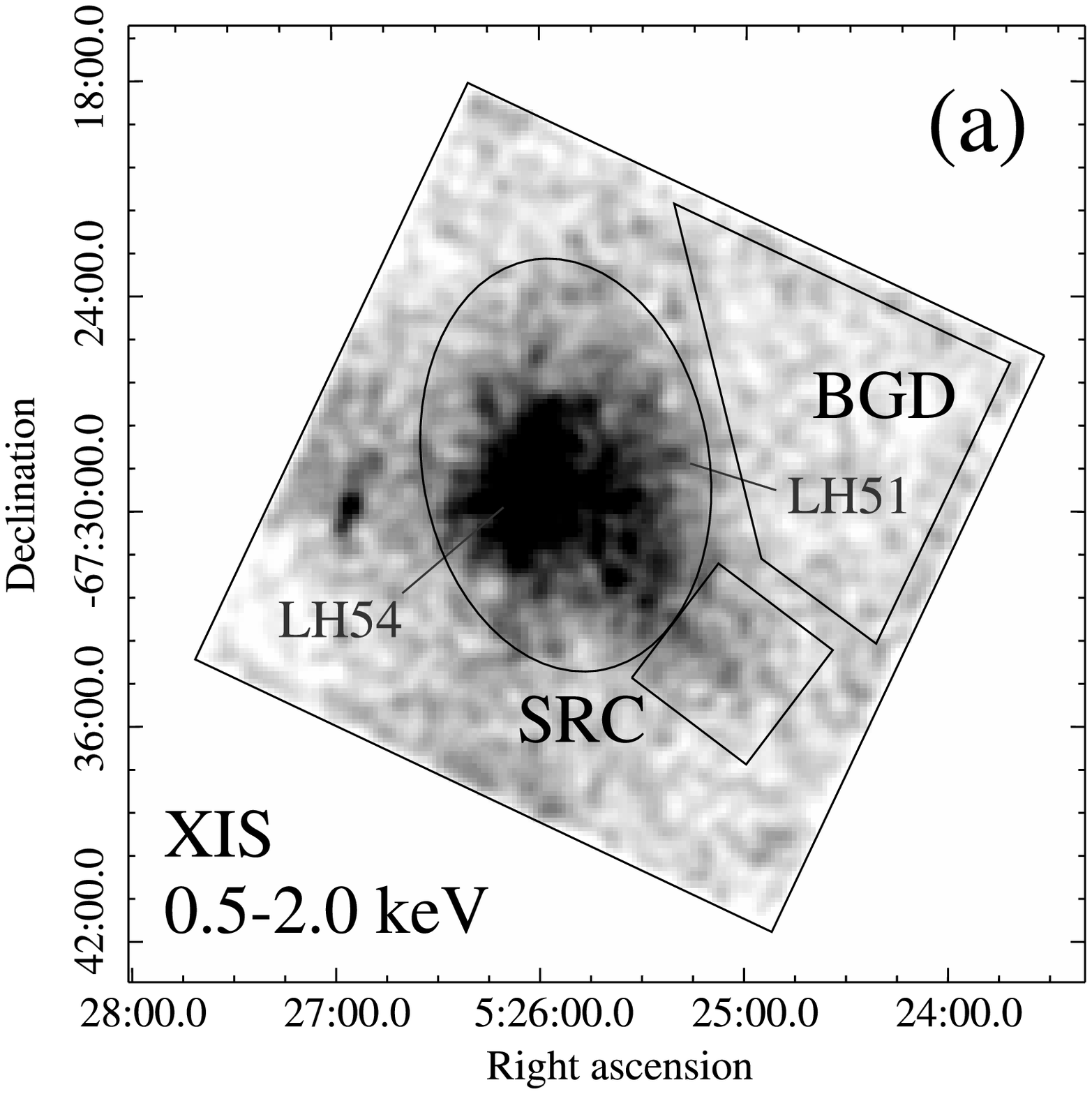}
  \includegraphics[scale=.34]{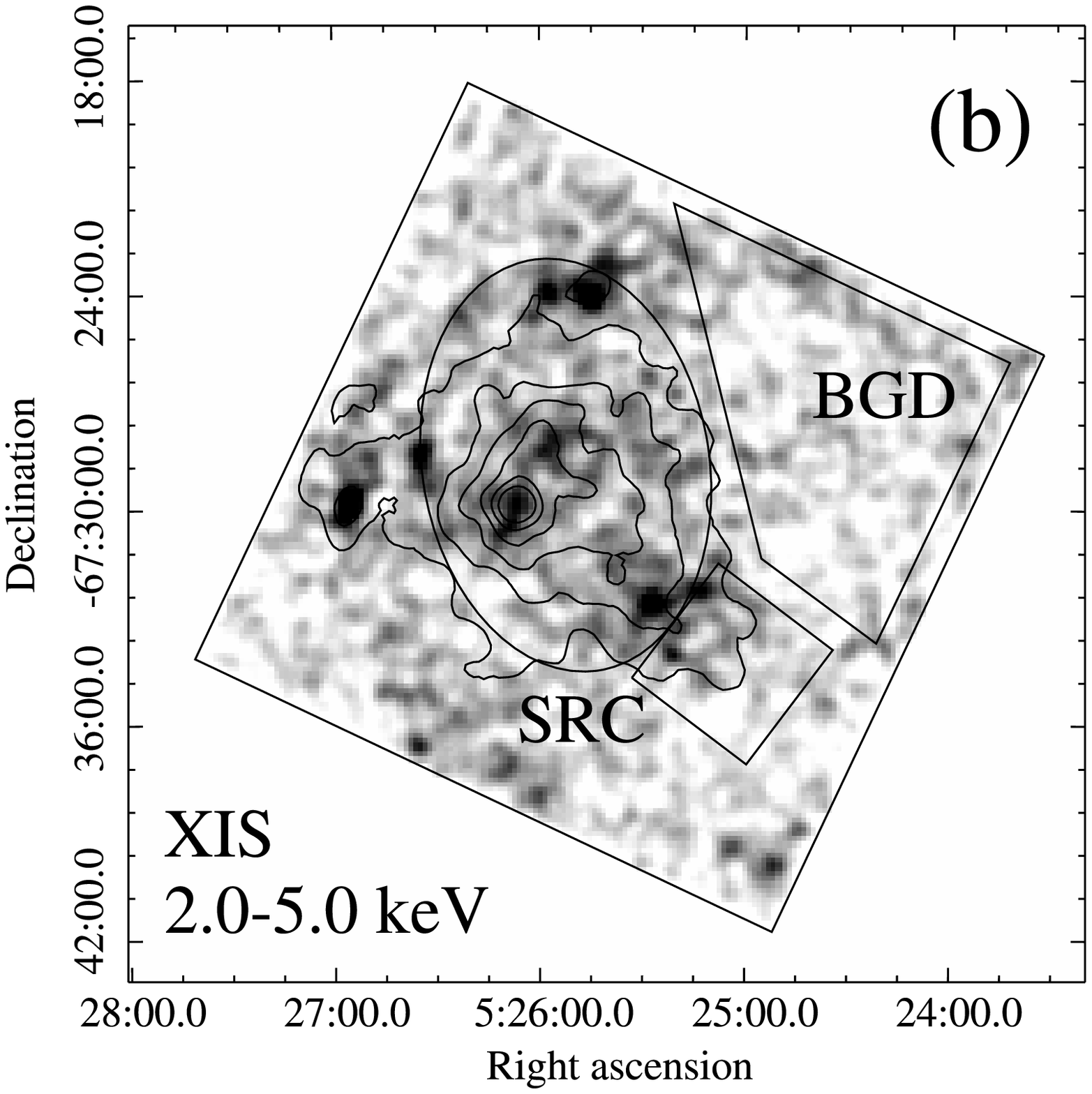}
  \includegraphics[scale=.31]{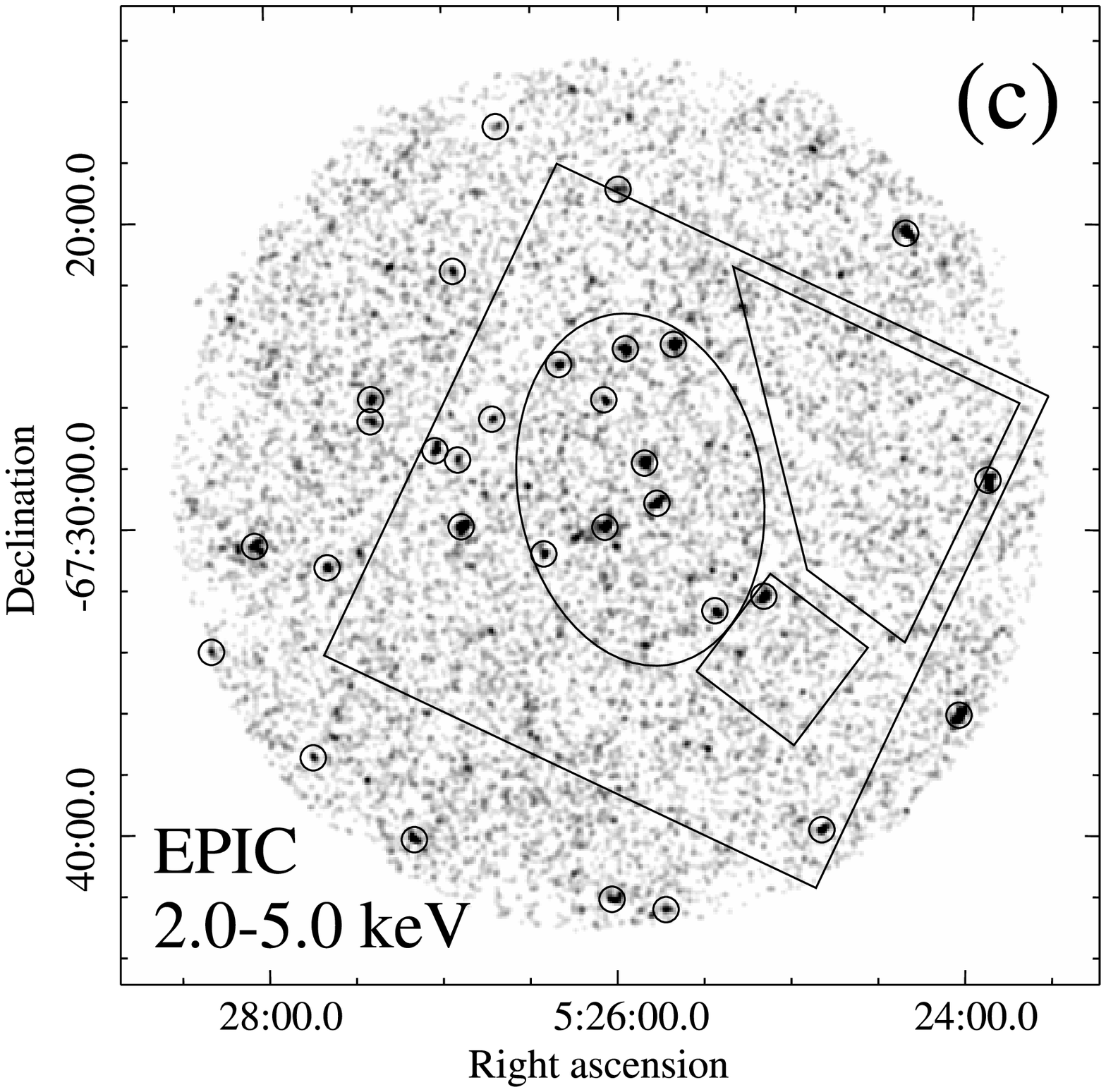}
\caption{Same as Figure~\ref{fig:n11_image}, but for the N51D region. 
  The ellipse and square are the source (SRC) regions for the diffuse 
  spectrum, whereas the polygon indicates the background (BGD) region. 
  \label{fig:n51d_image}}
  \end{center}
\end{figure*}

X-ray images of the N51D region are shown in Figure~\ref{fig:n51d_image} 
for the XIS soft X-ray band (a), the XIS hard X-ray band (b), and 
the EPIC hard X-ray band (c). 
The binning factors and smoothing Gaussian kernels are same as 
those used in Figure~\ref{fig:n11_image}. Similarly to N11, 
the diffuse structure was only observed in the soft-band image. 
The bright soft emission is coincident with the OB association LH54 
and is surrounded by the N51D \Ha\ shell 
(see also Figure~1 of Cooper et al.\ 2004).

The SRC region was selected to be the same as that in \cite{c04}: 
an ellipse with major and minor radii of $5.\!'8$ and $4'$, respectively,  
and a $4'$$\times$$4'$-square, indicated in Figure~\ref{fig:n51d_image}. 
Although a small circular region around the Wolf-Rayet star HD~36402 
was excluded from the spectral analysis in the previous work, 
here we included this region because no local structures were resolved 
in the low-resolution XIS image. 
The BGD region was selected to be an off-source polygonal region, 
but the CCD corner irradiated by the calibration sources was excluded.

Using the EPIC image (Figure~\ref{fig:n51d_image}c), we searched for 
hard discrete sources. As a result, ten sources were detected in 
the SRC region, while no point source was found in the BGD region. 
These sources were not clearly resolved in the XIS image. 
Therefore, we first analyzed the EPIC spectra of the point sources 
and then performed a model fit to the XIS data in the SRC region 
by adding the point-source component as a fixed model.

\subsection{XMM-Newton spectra of point-like sources}
\label{ssec:n51d_xmm}

\begin{figure}
  \begin{center}
    \rotatebox{270}{
    \includegraphics[scale=.31]{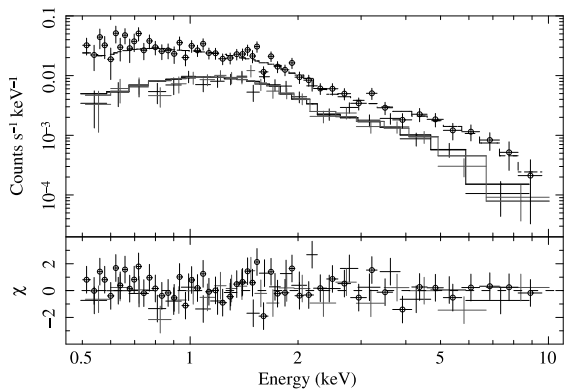}
    }
    \caption{Same as Figure~\ref{fig:n11_ptsrc}, but for N51D. 
      Ten point sources within the SRC region are integrated. 
  \label{fig:n51d_ptsrc}}
  \end{center}
\end{figure}

\begin{table}
\begin{center}
\caption{Flux distribution of the point sources within 
  the N51D SRC region.
  \label{tab:ptsrc_list}}
\begin{tabular}{cc}
\tableline\tableline
   Flux  &  Number  \\
  \tableline
  0.5--1.0  &  2  \\
  1.0--2.0  &  3  \\
  2.0--4.0  &  4  \\
  4.0--6.0  &  1  \\
  \tableline
\end{tabular}
\tablecomments{
  Observed fluxes are given in the spectral range of 2--10~keV, 
  and in unit of $10^{-14}$~\flux. 
}
\end{center}
\end{table}

EPIC spectra of the ten point sources in the SRC region were extracted 
and integrated. The background spectra were extracted from an annular 
region around each point source. The resultant background-subtracted 
spectra are shown in Figure~\ref{fig:n51d_ptsrc}. The spectra were well 
reproduced with an absorption power-law model (\chisq\ = 81/94). 
The best-fit absorption column and photon index were 
\NH\ = 1.3 (0.87--1.8) $\times 10^{21}$~cm$^{-2}$ and 
$\Gamma$ = 1.7 (1.6--1.8), respectively. The observed flux was obtained 
to be  2.4 (2.3--2.5) $\times 10^{-13}$~\flux\ in the 2--10~keV band. 
The flux distribution of the ten sources is given 
in Table~\ref{tab:ptsrc_list}.

\subsection{Suzaku spectra of extended emission}
\label{ssec:n51d_suzaku}

\begin{figure*}
  \begin{center}
    \rotatebox{270}{
    \includegraphics[scale=.31]{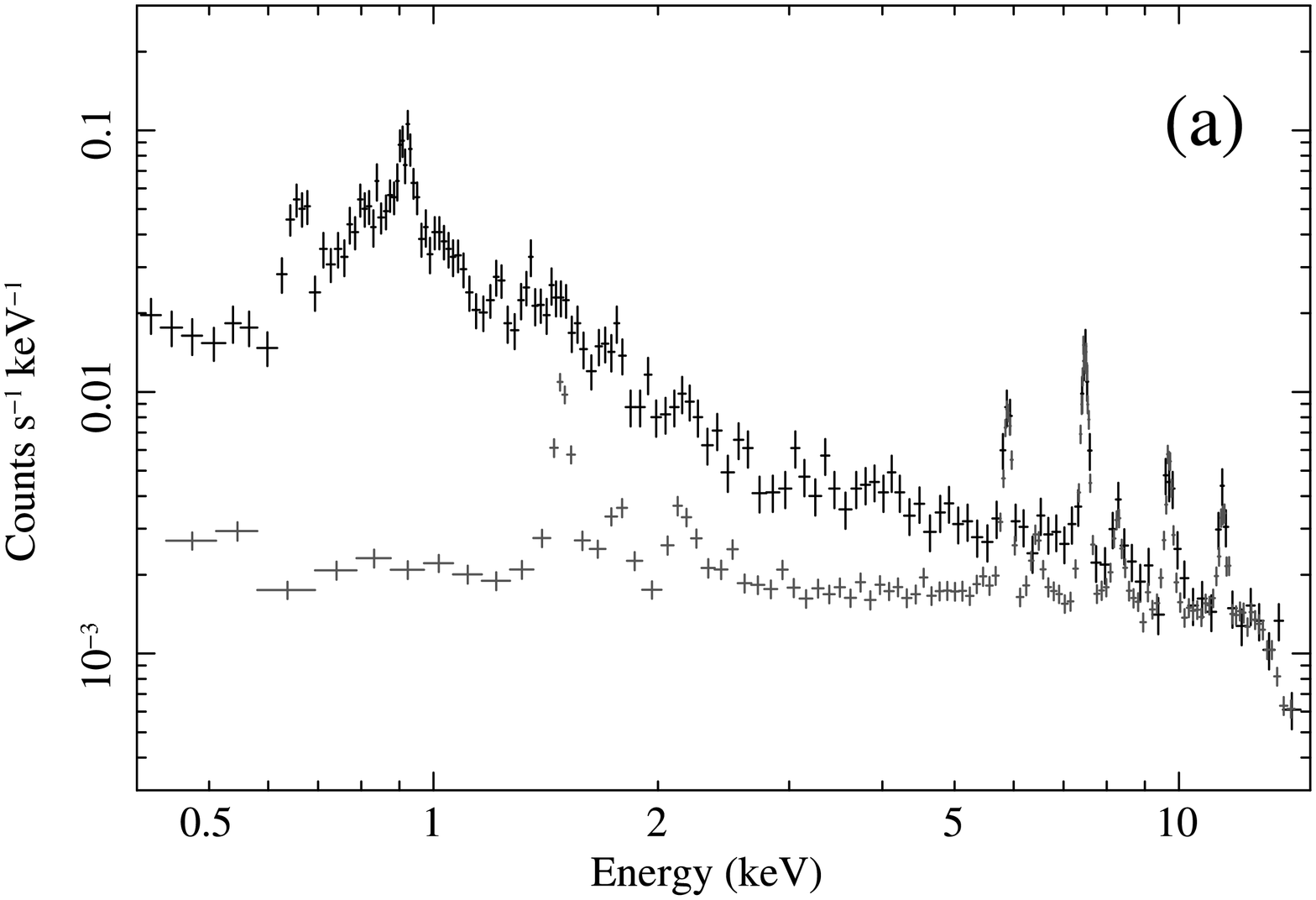}
    }
    \rotatebox{270}{
    \includegraphics[scale=.31]{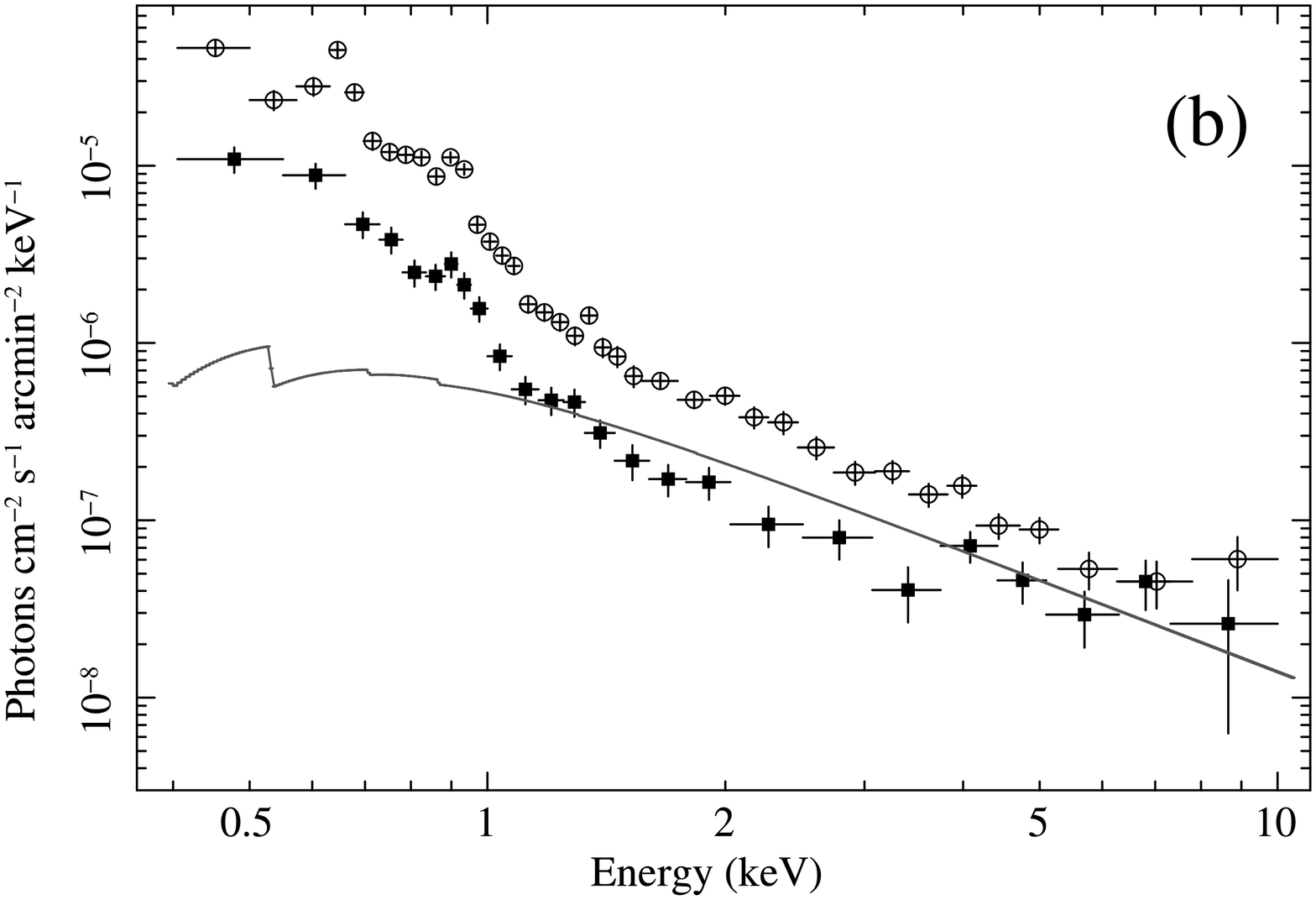}
    }
    \caption{Same as Figure~\ref{fig:n11_compare}, but for N51D. 
      The gray curve in the panel (b) indicates the spectral 
      model for the point sources derived by 
      the EPIC spectra (Figure~\ref{fig:n51d_ptsrc}). 
  \label{fig:n51d_compare}}
  \end{center}
\end{figure*}

\begin{figure}
  \begin{center}
    \rotatebox{270}{
    \includegraphics[scale=.32]{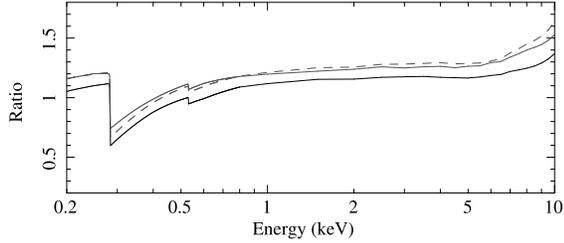}
    }
    \caption{Same as Figure~\ref{fig:n51d_ratio}, but for the 
      observation of N51D. 
  \label{fig:n51d_ratio}}
  \end{center}
\end{figure}

\begin{figure}
  \begin{center}
    \rotatebox{270}{
    \includegraphics[scale=.31]{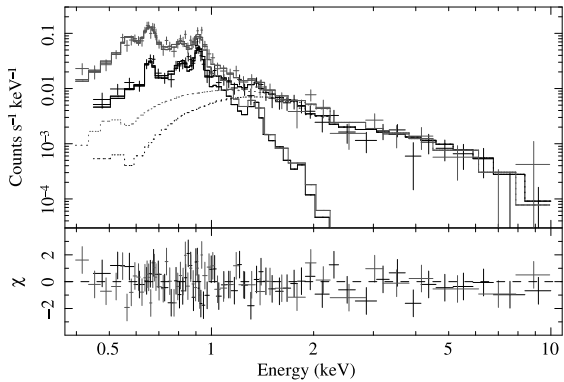}
    }
    \caption{XIS spectra of N51D, fitted with an APEC model 
      (solid lines). Black and gray represent FI and BI, 
      respectively. The dotted lines represent the spectra 
      of the point sources. 
  \label{fig:n51d_diffuse}}
  \end{center}
\end{figure}

\begin{table*}
\begin{center}
\caption{2--10~keV surfaces brightnesses of N51D region.
  \label{tab:sbn_n51d}}
\begin{tabular}{lccc}
  \tableline\tableline
  ~ & Solid angle & Flux [2--10~keV] & Surface brightness [2--10~keV] \\
  ~ & (arcmin$^2$) & ($\times 10^{-13}$~\flux) & ($\times 10^{-15}$~\sbn) \\
  \tableline
  SRC    &  88.9    & 5.2 (5.0--5.4)  &  5.9 (5.6--6.1) \\
  ~~(point sources) & ~ & 2.4 (2.3--2.5)  &  --- \\
  ~~(subtracted)    & ~ & 2.8 (2.5--3.0)  &  3.1 (2.9--3.4) \\
  BGD    &  48.4    & 1.7 (1.5--1.9)  &  3.5 (3.1--3.9) \\
  \tableline
\end{tabular}
\end{center}
\end{table*}

\begin{table}
\begin{center}
\caption{Best-fit spectral parameters for N51D.
  \label{tab:fit_n51d}}
\begin{tabular}{lc}
\tableline\tableline
  Parameter & value \\
  \tableline
  $N_{\rm H}^{\rm G}$ (cm$^{-2}$) & $6.0 \times 10^{20}$ (fixed) \\
  $N_{\rm H}^{\rm LMC}$ (cm$^{-2}$) & $< 5.2 \times 10^{20}$ \\
  $kT_e$ (keV) & 0.22 (0.20--0.23) \\
  O  (solar) &  0.14 (0.11--0.21)  \\
  Ne (solar) &  0.35 (0.26--0.49)  \\
  Fe (solar)  & 0.18 (0.12--0.27)  \\
  $n_e n_p V$ (cm$^{-3}$) & 9.8 (7.0--14) $\times ~10^{58}$ \\
  $F_{\rm X}^{\rm obs}$~(\flux) &  $5.6\times 10^{-13}$  \\
  $L_{\rm X}$~(\lumi) &  $2.4\times 10^{35}$  \\
  \tableline
  \chisq  & 92/103 = 0.89 \\
  \tableline
\end{tabular}
\end{center}
\end{table}

The spectral analysis of the {\it Suzaku} data was performed by 
essentially the same procedure as that described in 
Section~\ref{ssec:n11_suzaku}. 
An example of a raw spectrum in the SRC region and the corresponding 
NXB are compared in Figure~\ref{fig:n51d_compare}(a). 
Given that both spectra exhibit the same intensity of each fluorescent 
line, we considered that the NXB was successfully constructed. 
%The intensities of the Ni-\Ka\ fluorescent lines were obtained to 
%be 2.1 (1.7--2.4) $\times 10^{-3}$~counts~s$^{-1}$ for the SRC 
%and 2.3 (2.2--2.4) $\times 10^{-3}$~counts~s$^{-1}$ for the NXB, 
%fully consistent with each other. 
The NXB-subtracted spectra and the 2--10~keV surface brightnesses of 
the SRC and BGD regions were obtained as shown in 
Figure~\ref{fig:n51d_compare}(b) and Table~\ref{tab:sbn_n51d}, respectively. 
The SRC spectrum appears to show excess surface brightness in the hard 
X-ray band. However, upon subtracting the point-source flux from the SRC, 
we found that the remaining surface brightness is consistent with 
that in the BGD region.

We applied the vignetting correction to the BGD spectra and then 
subtracted them from the SRC spectra. The correction factors $f(E)$ 
were obtained as given in Figure~\ref{fig:n51d_ratio}. 
The resultant SRC spectra are shown in Figure~\ref{fig:n51d_diffuse}. 
The spectra of two active FIs were merged to improve the photon statistics. 
Several emission lines can be observed below $\sim$1~keV, while the 
spectrum in the hard band is featureless. 
Therefore, we fitted the spectrum with a VAPEC model plus the 
point-source component (a power-law) derived in Section~\ref{ssec:n51d_xmm}. 
The elemental abundances except for those of O, Ne, and Fe were fixed 
to the LMC values (Russell \& Dopita 1992; Hughes et al.\ 1998). 
We assumed the Galactic absorption column to be 
$N_{\rm H}^{\rm G}$ = $6.0\times 10^{20}$~cm$^{-2}$, 
in accordance with \cite{dl}. 
The fit was acceptable with \chisq\ = 92/103. 
The best-fit parameters and models are respectively shown in 
Table~\ref{tab:fit_n51d} and in Figure~\ref{fig:n51d_diffuse} 
with solid and dotted lines. 
Note that the point-source component was not fitted here. 
The model successfully reproduced the spectral shape as well as 
the flux of the hard X-rays, and hence no additional nonthermal 
component was required to improve the fit.

\cite{c04} argued that the EPIC spectrum of the diffuse emission 
exhibited thermal and nonthermal components, and the former accounted 
for $\sim$70\% of the total observed flux (in the 0.3--3.0~keV band) 
of $1.1\times 10^{-12}$~\flux. 
With the given temperature and absorption column, the flux of the 
thermal emission was calculated to be $\sim$$6 \times 10^{-13}$~\flux 
in the 0.5--2.0~keV band. This is comparable to our result.

The enhancement of O and/or Ne abundances was reported 
in the works of \cite{bomans03} and \cite{c04}. 
On the other hand, our spectrum does not show any evidence of 
a significant deviation from the typical LMC abundances, suggesting 
that the hot gas mainly originates from ISM with only slight enrichment 
by SN ejecta.

\section{Discussion}
\label{sec:discussion}

\begin{figure*}
  \begin{center}
    \rotatebox{270}{
    \includegraphics[scale=.31]{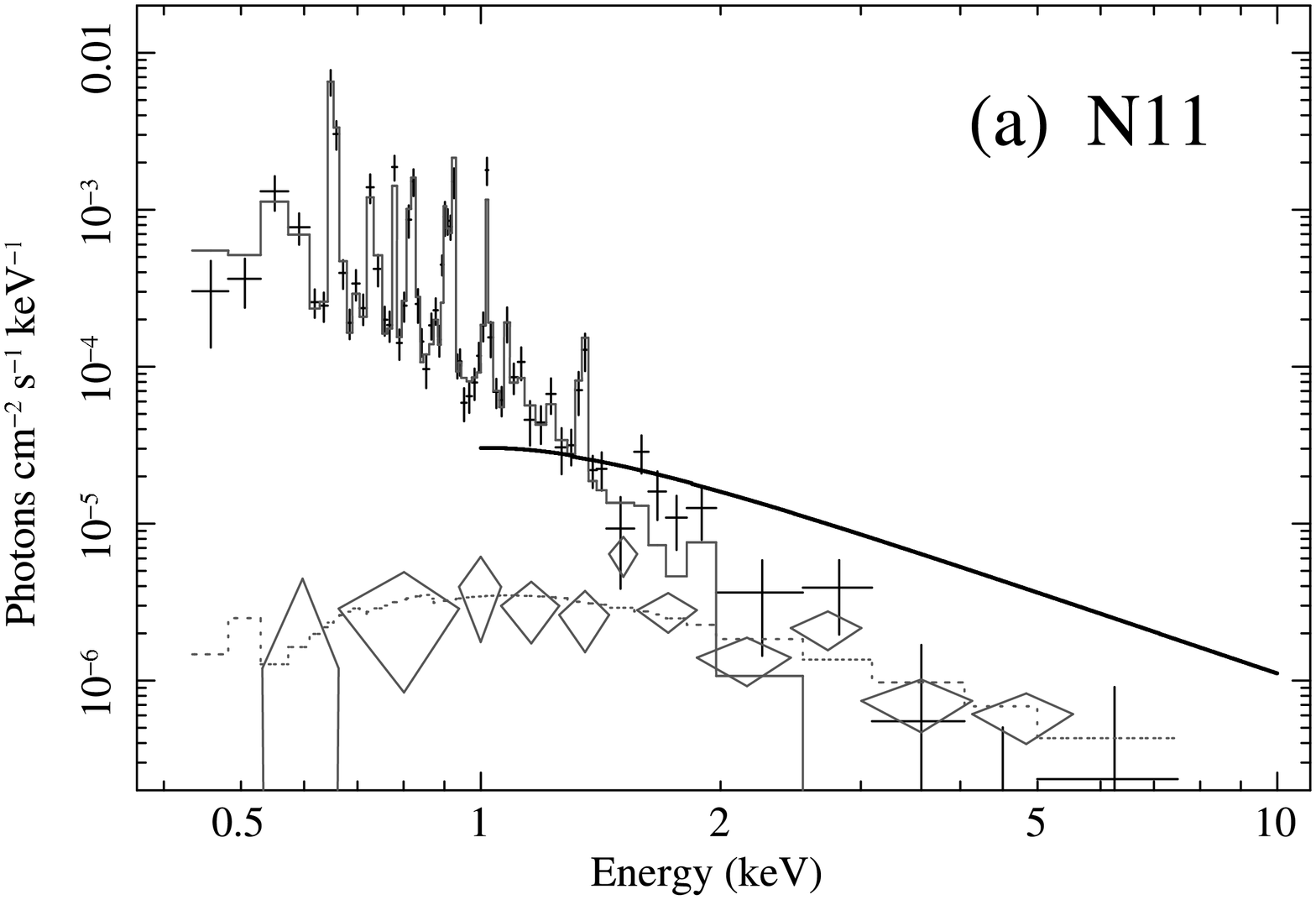}
    }
    \rotatebox{270}{
    \includegraphics[scale=.31]{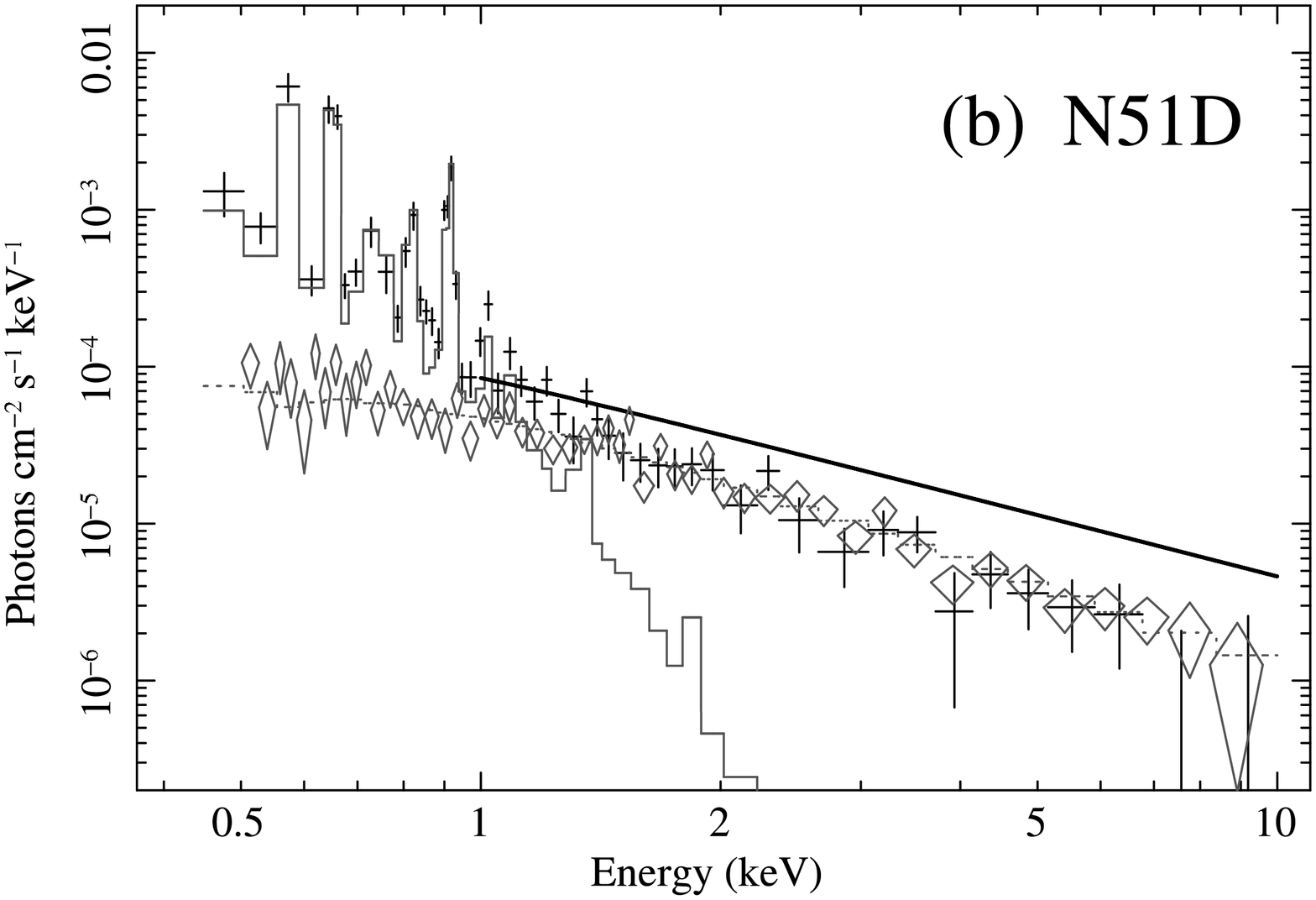}
    }
    \caption{XIS-FI unfolded spectra (black crosses) of N11 (a) 
      and N51D (b). The EPIC-pn spectra of the point sources 
      are also shown with gray diamonds. Solid and dashed gray lines 
      are the model components of the diffuse and point sources, 
      respectively. The spectra of ``nonthermal emission'' claimed 
      by \cite{m09} and \cite{c04} (for N11 and N51D, respectively) 
      are indicated with the black thick lines. 
      The data points in the energies above $\sim 2$~keV are obviously 
      far below the claimed fluxes although the point 
      sources are not subtracted from the XIS spectra. 
  \label{fig:unfold}}
  \end{center}
\end{figure*}

The X-ray spectra of the SBs N11 and N51D are well reproduced by 
single-component thin-thermal plasmas (with the models for the 
group of point sources). 
Figure~\ref{fig:unfold} shows the unfolded spectra of both SBs. 
The presence of additional nonthermal emission, argued in previous 
works (N11: Maddox et al.\ 2009; N51D: Cooper et al.\ 2004), 
is much less evident in our detailed analyses, supporting 
the earlier results by \cite{naze04} and \cite{bomans03}. 
The 2--10~keV flux of the diffuse component 
$F_{\rm D}$ ($\pm~\sigma_{\rm D}$) for each SB can be calculated as 
\begin{eqnarray}
  F_{\rm D} &=& \left[ F_{\rm S} -  F_{\rm PS} - 
    \frac{\Omega_{\rm S}}{\Omega_{\rm B}} \cdot
    F_{\rm B} \right] \pm \sigma_{\rm D}  \\
  \sigma_{\rm D} &=& \left[ \sigma_{\rm S}^2 +  \sigma_{\rm PS}^2 + 
    \left( \frac{\Omega _{\rm S}}{\Omega _{\rm B}} \right)^2  
    \sigma_{\rm B}^2 \right]^{1/2},
\end{eqnarray}
where $F_{\rm S}$, $F_{\rm B}$, and $F_{\rm PS}$ 
(and $\sigma_{\rm S}$, $\sigma_{\rm B}$, $\sigma_{\rm PS}$) are 
the 2--10~keV fluxes (and uncertainties) of the SRC and BGD regions 
and that of the point sources.
Using the values given in Tables~\ref{tab:sbn_n11} and \ref{tab:sbn_n51d}, 
we obtain 3$\sigma$ upper limits of the fluxes to be 
$3.6 \times 10^{-14}$~\flux\ and $4.7 \times 10^{-14}$~\flux\ 
for N11 and N51D, respectively.

For N11, \cite{m09} claimed that the hard X-rays, detected at energies of 
up to $\sim$7~keV, required a power-law component with a photon index of 
$\Gamma \sim 1.7$ and an observed flux of $\sim$$3.0\times 10^{-13}$~\flux\ 
(in the 0.2--10~keV band)\footnote{The energy range from which the flux was 
derived is not given in \cite{m09}. Therefore, we independently estimated 
it from reported values of the absorption column density, photon index, 
and the ratio of the observed to unabsorbed fluxes.}$\!\!$. 
For N51D, the photon index and observed flux in the 0.3--3.0~keV band 
were reported to be $\Gamma \sim 1.3$ and $\sim$$3.3\times 10^{-13}$~\flux, 
respectively \citep{c04}. 
To directly compare these values with our results, these fluxes are 
modified to values in the 2--10~keV band, 
giving $2.3\times 10^{-13}$~\flux\ for N11 and 
$7.1\times 10^{-13}$~\flux\ for N51D. These values are about 
one order higher than the upper limits that we have obtained. 
The spectral models for the claimed hard X-ray emissions are shown 
in Figure~\ref{fig:unfold} with black solid lines. 
Note that the point sources had been excluded from the EPIC spectrum 
of N51D \citep{c04}. 
Nevertheless, both models give fluxes clearly exceeding those given 
by the XIS spectrum in which the point-source fluxes are contained.

Given that the point-source flux in N11 is negligibly small 
(Fig.~\ref{fig:unfold}), the detection of hard X-rays claimed in the 
previous report was probably due to incorrect background subtraction. 
In fact, \cite{m09} commented that the subtraction of 
Ni lines originating from the NXB was unsuccessful. 
The NXB spectrum of the XIS consists of emission-line and continuum 
components, and their intensities are correlated with each other 
\citep{tawa08}. Therefore, this also caused the incomplete subtraction 
of the continuum component in the previous reports. 
In our analyses, on the other hand, the fluorescent NXB lines 
are accurately subtracted (see Section~\ref{sssec:n11_suzaku1}), 
and hence the background reduction is considered to be successful. 
The problem concerning N51D is analogous to the case of N11, 
because the residual of the Al-\Ka\ line due to incomplete 
background subtraction has been reported \citep{c04}. 
We point out that the analysis of faint extended sources, such as 
SBs in the LMC, is very sensitive to the estimation of the NXB and 
the vignetting effect, and thus should be performed with great care.

\begin{table}
\begin{center}
\caption{Comparison of nonthermal luminosities with other sources. 
  \label{tab:comparison}}
\begin{tabular}{lcc}
\tableline\tableline
  Object &  $L_{\rm X}$~(\lumi)\tablenotemark{a}  &  Reference  \\
  \tableline
  N11  & $<$ $1 \times 10^{34}$  & This work. \\
  N51D & $<$ $1 \times 10^{34}$  & This work. \\
  30~Dor~C & $2 \times 10^{35}$ & Yamaguchi et al.\ (2009) \\
  Arches & $1 \times 10^{34}$ & Tsujimoto et al.\ (2007) \\
  NGC~6334 & $8 \times 10^{32}$ & Ezoe et al.\ (2006) \\
  RCW38 & $5 \times 10^{32}$ & Wolk et al.\ (2002) \\
  Westerlund~1 & $3 \times 10^{34}$ & Muno et al.\ (2006) \\
  Westerlund~2 & $2 \times 10^{34}$ & Fujita et al.\ (2009) \\
  \tableline
\end{tabular}
\tablecomments{
  $^{a}$Luminosity in the 2--10~keV band. 
  }
\end{center}
\end{table}

Nonthermal X-ray emissions have so far been discovered from several 
Galactic star-forming regions, the Arches cluster (Tsujimoto et al.\ 2007), 
NGC~6334 (Ezoe et al.\ 2006), RCW~38 (Wolk et al.\ 2002), 
Westerlund~1 (Muno et al.\ 2006), and Westerlund~2 (Fujita et al.\ 2009). 
In these cases, however, the emissions are correlated with relatively 
compact (1--10-pc-scale) regions around an OB association or in a 
molecular cloud core. 30~Dor~C is another example of a star-forming 
region that exhibits nonthermal (synchrotron) X-rays \citep{bamba04}. 
However, the X-ray shell
can be interpreted as a single middle-aged SNR expanding rapidly 
inside the SB cavity \citep{yama09}. 
Therefore, at present, there is no evidence for large-scale 
($\sim$100~pc) nonthermal emission distributed entirely in an SB. 
As given in Table~\ref{tab:comparison}, the luminosities of the 
nonthermal X-ray emissions in N11 and N51D are, if any, less than 
those of 30~Dor~C and some of the other star-forming regions.

It is theoretically expected that efficient cosmic-ray acceleration can 
take place in SBs due to repeated SN shocks and/or magnetic turbulence 
(e.g., Bykov \& Fleishman 1992; Parizot et al.\ 2004). 
Moreover, Butt \& Bykov (2008) suggested that more than 10\% of the 
energy stored in an SB can be constantly transferred to accelerating 
cosmic-ray particles at the evolution stage of N11 and N51D (a few Myr). 
If this is the case, accelerated electrons would quickly lose their 
energies via radiation so that no nonthermal X-rays would be detected. 
It is still an open issue whether or not SBs can efficiently 
accelerate cosmic-rays, but this is beyond the scope of this paper. 
Future detailed observations and the development of theoretical models 
are required to solve this problem.

\newpage

\section{Summary}
\label{sec:summary}

We have analyzed the {\it Suzaku} XIS data of the extended X-ray 
emission from the SBs N11 and N51D. 
The analysis of the {\it XMM-Newton} EPIC data has also been 
performed to quantify the contribution from the point-like sources. 
In both SBs, the spectra of the diffuse components are well 
represented by thin thermal plasma at collisional ionization 
equilibrium. The electron temperatures derived for N11 and N51D are 
$\sim$0.3~keV and $\sim$0.2~keV, respectively. 
The elemental abundances are comparable with average LMC values, 
suggesting that the thermal emission mainly originates from the 
shock-heated ISM with only slight SN enrichment.

Neither N11 nor N51D show any evidence for hard X-ray emission, 
contrary to the previous claims by \cite{m09} and \cite{c04}. 
The 3$\sigma$ upper limits of the 2--10~keV flux are 
$3.6 \times 10^{-14}$~\flux\ for N11 and 
$4.7 \times 10^{-14}$~\flux\ for N51D.  
The published claims of the detection of nonthermal X-rays are 
thought to be simply due to inadequate background subtraction. 
Careful analyses (i.e., NXB and point-source subtraction, 
vignetting correction, and so on) are required to obtain 
an accurate flux for faint extended sources, such as N11 and N51D. 
As the result of this work, no observational evidence for 
the nonthermal X-ray emission widely associated with SBs remains, 
with the exception of 30~Dor~C (however, this is likely to be 
a single remnant of a recent SN explosion). 
It is still an open issue whether or not SBs can efficiently 
accelerate cosmic-rays. Observations of more SBs with longer exposures 
as well as the development of theoretical calculations are required.

\acknowledgments

We thank Yoshihiro Ueda for useful comments on the CXB. 
H.Y.\ is supported by the Special Postdoctoral Researchers Program 
in RIKEN, and the Grant-in-Aid for Young Scientists from the Ministry of 
Education, Culture, Sports, Science and Technology (MEXT) of Japan.
M.S.\ is a Research Fellow of Japan Society for the Promotion of 
Science (JSPS).


\begin{thebibliography}{}

\bibitem[Anders \& Grevesse(1989)]{angr} Anders, E., \& 
Grevesse, N.\ 1989, \gca, 53, 197 


\bibitem[Bamba et al.(2004)]{bamba04} Bamba, A., Ueno, M., 
Nakajima, H., \& Koyama, K.\ 2004, \apj, 602, 257 

\bibitem[Bomans et al.(2003)]{bomans03} Bomans, D.~J., 
Rossa, J., Weis, K., \& Dennerl, K.\ 2003, 
A Massive Star Odyssey: From Main Sequence to Supernova, 212, 637 

\bibitem[Butt \& Bykov(2008)]{2008ApJ...677L..21B} Butt, Y.~M., \& 
Bykov, A.~M.\ 2008, \apjl, 677, L21 

\bibitem[Bykov \& Fleishman(1992)]{1992MNRAS.255..269B} Bykov, A.~M., \& 
Fleishman, G.~D.\ 1992, \mnras, 255, 269 


\bibitem[Chu \& Mac Low(1990)]{1990ApJ...365..510C} Chu, Y.-H., \& 
Mac Low, M.-M.\ 1990, \apj, 365, 510 

\bibitem[Cooper et al.(2004)]{c04} Cooper, R.~L., Guerrero, 
M.~A., Chu, Y.-H., Chen, C.-H.~R., \& Dunne, B.~C.\ 2004, \apj, 605, 751 

\bibitem[Dickey \& Lockman(1990)]{dl} Dickey, J.~M., 
\& Lockman, F.~J.\ 1990, \araa, 28, 215 

\bibitem[Dunne et al.(2001)]{dunne01} Dunne, B.~C., Points, 
S.~D., \& Chu, Y.-H.\ 2001, \apjs, 136, 119 

\bibitem[Ezoe et al.(2006)]{2006ApJ...638..860E} Ezoe, Y., Kokubun, M., 
Makishima, K., Sekimoto, Y., \& Matsuzaki, K.\ 2006, \apj, 638, 860 

\bibitem[Fujita et al.(2009)]{2009PASJ...61.1229F} Fujita, Y., 
Hayashida, K., Takahashi, H., \& Takahara, F.\ 2009, \pasj, 61, 1229 

\bibitem[Hatano et al.(2006)]{2006AJ....132.2653H} Hatano, H., et al.\ 
2006, \aj, 132, 2653 

\bibitem[Henize(1956)]{henize56} Henize, K.~G.\ 1956, \apjs, 2, 
315 

\bibitem[Hughes et al.(1998)]{1998ApJ...505..732H} Hughes, J.~P., Hayashi, 
I., \& Koyama, K.\ 1998, \apj, 505, 732 

\bibitem[Koyama et al.(2007)]{2007PASJ...59S..23K} Koyama, K., et al.\ 
2007, \pasj, 59, S23 

\bibitem[Lucke \& Hodge(1970)]{lh} Lucke, P.~B., \& 
Hodge, P.~W.\ 1970, \aj, 75, 171 

\bibitem[Mac Low et al.(1998)]{mac98} Mac Low, M.-M., Chang, 
T.~H., Chu, Y.-H., Points, S.~D., Smith, R.~C., 
\& Wakker, B.~P.\ 1998, \apj, 493, 260 

\bibitem[Mac Low \& McCray(1988)]{mac88} Mac Low, M.-M., \& 
McCray, R.\ 1988, \apj, 324, 776 

\bibitem[Maddox et al.(2009)]{m09} Maddox, L.~A., Williams, 
R.~M., Dunne, B.~C., \& Chu, Y.-H.\ 2009, \apj, 699, 911 

\bibitem[McCray \& Snow(1979)]{1979ARA&A..17..213M} McCray, R., \& 
Snow, T.~P., Jr.\ 1979, \araa, 17, 213 

\bibitem[Mitsuda et al.(2007)]{2007PASJ...59S...1M} Mitsuda, K., et al.\ 
2007, \pasj, 59, S1 

\bibitem[Morrison \& McCammon(1983)]{1983ApJ...270..119M} Morrison, R., \& 
McCammon, D.\ 1983, \apj, 270, 119 

\bibitem[Muno et al.(2006)]{2006ApJ...650..203M} Muno, M.~P., Law, C., 
Clark, J.~S., Dougherty, S.~M., de Grijs, R., Portegies Zwart, S., 
\& Yusef-Zadeh, F.\ 2006, \apj, 650, 203 

\bibitem[Naz{\'e} et al.(2004)]{naze04} 
Naz{\'e}, Y., Antokhin, I.~I., Rauw, G., Chu, Y.-H., Gosset, E., \& 
Vreux, J.-M.\ 2004, \aap, 418, 841 

\bibitem[Oey \& Smedley(1998)]{1998AJ....116.1263O} Oey, M.~S., \& 
Smedley, S.~A.\ 1998, \aj, 116, 1263 

\bibitem[Parizot et al.(2004)]{2004A&A...424..747P} Parizot, E., 
Marcowith, A., van der Swaluw, E., Bykov, A.~M., \& Tatischeff, V.\ 
2004, \aap, 424, 747 

\bibitem[Persson et al.(2004)]{2004AJ....128.2239P} Persson, S.~E., Madore, 
B.~F., Krzemi{\'n}ski, W., Freedman, W.~L., Roth, M., 
\& Murphy, D.~C.\ 2004, \aj, 128, 2239 

\bibitem[Russell \& Dopita(1992)]{rd} Russell, S.~C., 
\& Dopita, M.~A.\ 1992, \apj, 384, 508 

\bibitem[Serlemitsos et al.(2007)]{2007PASJ...59S...9S} Serlemitsos, P.~J., 
et al.\ 2007, \pasj, 59, S9 

\bibitem[Smith et al.(2001)]{2001ApJ...556L..91S} Smith, R.~K., Brickhouse, 
N.~S., Liedahl, D.~A., \& Raymond, J.~C.\ 2001, \apj, 556, L91 

\bibitem[Str{\" u}der et al.(2001)]{2001A&A...365L..18S} Str{\" u}der, 
L.~et al.\ 2001, \aap, 365, L18 

\bibitem[Tawa et al.(2008)]{tawa08} Tawa, N., et al.\ 2008, 
\pasj, 60, S11

\bibitem[Tsujimoto et al.(2007)]{2007PASJ...59S.229T} Tsujimoto, M., Hyodo, 
Y., \& Koyama, K.\ 2007, \pasj, 59, 229 

\bibitem[Turner et al.(2001)]{2001A&A...365L..27T} Turner, M.~J.~L.~et al.\ 
2001, \aap, 365, L27 

\bibitem[Walborn et al.(1999)]{1999AJ....118.1684W} Walborn, N.~R., 
Drissen, L., Parker, J.~W., Saha, A., MacKenty, J.~W., 
\& White, R.~L.\ 1999, \aj, 118, 1684 

\bibitem[Walborn \& Parker(1992)]{1992ApJ...399L..87W} Walborn, N.~R., \& 
Parker, J.~W.\ 1992, \apjl, 399, L87 

\bibitem[Weaver et al.(1977)]{1977ApJ...218..377W} Weaver, R., McCray, R., 
Castor, J., Shapiro, P., \& Moore, R.\ 1977, \apj, 218, 377 

\bibitem[Wolk et al.(2002)]{2002ApJ...580L.161W} Wolk, S.~J., Bourke, 
T.~L., Smith, R.~K., Spitzbart, B., \& Alves, J.\ 2002, \apjl, 580, L1

\bibitem[Yamaguchi et al.(2009)]{yama09} Yamaguchi, H., Bamba, 
A., \& Koyama, K.\ 2009, \pasj, 61, 175 




\end{thebibliography}
\end{document}